\pdfoutput=1
\documentclass[acmsmall,screen,nonacm]{acmart}
\usepackage{newunicodechar}
\newunicodechar{−}{\ensuremath{-}}
\newunicodechar{∩}{\ensuremath{\cap}}
\newunicodechar{κ}{\ensuremath{\kappa}}
\newunicodechar{σ}{\ensuremath{\sigma}}
\newunicodechar{μ}{\ensuremath{\mu}}
\newunicodechar{π}{\ensuremath{\pi}}
\newunicodechar{θ}{\ensuremath{\theta}}
\newunicodechar{∈}{\ensuremath{\in}}
\newunicodechar{∖}{\ensuremath{\setminus}}
\newunicodechar{≈}{\ensuremath{\approx}}
\newunicodechar{′}{\ensuremath{'}}
\newunicodechar{±}{\ensuremath{\pm}}
\newunicodechar{×}{\ensuremath{\times}}
\newunicodechar{·}{\ensuremath{\cdot}}
\newunicodechar{≤}{\ensuremath{\le}}
\newunicodechar{≥}{\ensuremath{\ge}}
\newunicodechar{∎}{\ensuremath{\qed}}
\newunicodechar{∅}{\ensuremath{\emptyset}}
\usepackage{longtable,array,calc,booktabs,graphicx,colortbl}
\usepackage{etoolbox}
\AtBeginEnvironment{longtable}{\footnotesize\setlength{\tabcolsep}{3pt}}
\usepackage{xurl}
\usepackage{needspace}
\AtBeginEnvironment{thebibliography}{\raggedright}
\setkeys{Gin}{width=\linewidth,keepaspectratio}
\providecommand{\tightlist}{\setlength{\itemsep}{0pt}\setlength{\parskip}{0pt}}

\setcopyright{none}
\renewcommand\footnotetextcopyrightpermission[1]{}
\begin{document}
\title{How Fast Stored Household Assignments Go Stale: Evidence from North Carolina Voter Rolls}
\author{Prashant Chaudhary}
\orcid{0009-0001-1980-109X}
\affiliation{\institution{Independent Researcher}\city{Santa Clara}\state{CA}\postcode{95051}\country{USA}}
\begin{abstract}
Household files and identity graphs are judged by whether they still
find the person. That test misses what they store. A household
assignment can go stale while every person-level link in it stays right,
because the people around the person change. We measure how fast, for
households derived from a voter register, by following North Carolina
registrants on a persistent identifier across 40 distinct snapshots from
November 2008 to November 2026 (3.6 million clients at the 2016 build in
60 counties). Eight years after the 2016 build, the graph finds the
address of 0.831 of registrants still listed, yet its household is
exactly right for only 0.322 of those found (0.309 reading stale
listings as departures). A mover's stored household is right about one
time in five at every build from two to eight years, so the loss runs
through who stays and how households that stay change. Stricter
thresholds of a probabilistic name-and-address matcher raise accuracy
only to about the stayers' currency, mainly by selecting stayers. Lookup
tables of currency by build-time features cut Brier score by 16\% to
19\% against evidence age alone and 6\% to 10\% against evidence age and
household size, on held-out counties and a later file. A multistate
model of household change does not beat a table with the same inputs,
and the tables choose which quarter of a file to refresh no better than
evidence age and household size do. In Ohio and Washington at eight to
nine years the tables keep their ranking after an in-sample intercept
shift; the accounting's level holds in Washington and not in Ohio.
Simulated build error of 5\% to 21\% costs 2\% to 10\% of eight-year
currency.
\end{abstract}
\ccsdesc[500]{Information systems~Entity resolution}
\ccsdesc[300]{Information systems~Data cleaning}
\keywords{data quality, data currency, household composition, record linkage, identity graphs, voter registration records}
\maketitle
\hypertarget{introduction}{%
\needspace{4\baselineskip}
\section{Introduction}\label{introduction}}

Suppose a file stores, for every person, the people that person lives with, and suppose the evidence behind it is eight years old. The system can still find most of the people. Is the household it hands back still the household?

Two communities store households. Vendors sell household files and identity graphs. Researchers and statistical offices build household keys from administrative addresses and attach them to people for years. Both judge the result by match rate, the share of people the system can still find. Match rate says nothing about the people around the person, and they change. Members leave, die, come of age or move in around someone who never moves. This paper measures how long a stored household stays right, and what a system that answers from old evidence returns once it does not: the right household, a wrong one, or nothing.

North Carolina's voter rolls let us answer for the household a register records. Take a household graph built from the 2016 register, with 3.6 million clients in 60 counties; a client is a person whose household the system is asked for. Eight years later the graph still finds, by address, 83.1\% of the clients who are still registered. The registered household it returns is exactly right for 32.2\% of those it finds (30.9\% if the register's stale listings are read as departures one Election Day ahead, section 6.1). By match rate the graph still works. By the households it returns, about two answers in three are no longer exactly right. Rebuild the graph from evidence ten months old and it is right for 75.5\% to 85.8\% of its matches across scoring dates (80.0\% at eight years; 69.3\% under the same reading).

Three literatures come close to this question. Data-quality research defines the currency of a stored value as the probability that it is still valid given its age. It estimates that probability for one attribute of one entity from a fitted or assumed lifetime distribution {[}42, 44{]}, and it has evaluated such estimates against observed validity {[}43{]}. Linkage research treats match rate as one side of a tradeoff against false links {[}1, 8, 18{]}. Demography has measured how household composition changes, from person-level panels and from linked annual registers that combine an electoral roll with consumer data {[}57, 58, 84{]}. A stored household assignment sits between these. It is a group-valued record, and we found no study that measures its currency as a system stores and returns it: anchored on an address or a person, scored against a reference that follows persons by a persistent identifier, and broken down by what the system answers.

The systems we study answer by finding an \textbf{anchor} in their evidence and returning the household recorded around it at the build date. An address anchor resolves the client's current address and returns whoever was recorded there. A person anchor finds the client's own build-time record and returns the household recorded around it. Match rate measures whether the anchor can still be found. Accuracy depends on whether the household around the anchor is still the household at the scoring date. We call that its \textbf{currency}. For the 2016 build at two years, the address is found for 0.954 of registrants still listed and its household is right for 0.701 of those address matches; the eight-year values are those above.

We make four contributions:

\begin{enumerate}
\def\labelenumi{\arabic{enumi}.}
\tightlist
\item
  \textbf{A longitudinal reference for group-valued records, and currency scored against it.} We follow the registered co-residence of North Carolina registrants across 41 snapshot files (40 distinct extracts) by a persistent identifier, with no matcher. Exits are classified by the register's removal reasons, and its stale listings are read five ways. On that reference we score currency, the probability that the household around a system's anchor is still exactly right, for registrants in trackable households, with three outcomes per client (sections 3, 4 and 6).
\item
  \textbf{The currency function and where its loss comes from.} A mover's stored household is right about one time in five at every build from two to eight years. The loss therefore runs through the share of clients who stay and through change inside households that stay. Currency falls with evidence age under every reading of stale listings. The accounting is reproduced descriptively on held-out counties, and it meets its level and decomposition tolerances in Washington's register but not in Ohio's (sections 5, 6 and 9).
\item
  \textbf{What a system can and cannot do about it.} Tightening the threshold of the probabilistic name-and-address matcher we measure raises the exact share of its matches only to about the stayers' currency of the population it answers (equation (3)). This is the currency form of the selection that linkage rules impose {[}86{]}. On the clients it matches, replacing its person links with perfect ones changes the exact share by about 0.001, because many of the matcher's wrong links return the client's own build household. An annual rebuild raises the exact share far more than any threshold does. Simulated build error costs less currency than its face value at long ages (sections 6.4 and 7).
\item
  \textbf{Predictions of currency tested on held-out data.} Lookup tables of currency by build-time features, the supplemental-data design of Heinrich and Klier {[}41{]} with a set-valued target, reduce Brier score against evidence age alone and against evidence age and household size, on held-out counties and on a later file, including its counties outside development. A multistate model of household change derives currency and its stayer and mover split from one-step transitions, as Zak and Even {[}96{]} do for one attribute. It beats the coarser table and trails the table with the same co-member ages, and its derived mover currency misses the measured one. When only a quarter of the file can be refreshed, neither table met its criterion against evidence age and household size (sections 8 and 9).
\end{enumerate}

\textbf{Method.} North Carolina gives each registrant a persistent identifier (NCID). Following NCIDs across 41 public snapshot files from November 2008 to November 2026 (40 distinct extracts) shows, for each registrant with one listing at a date, who shared their recorded residential unit and when each record changed or left the rolls. Our simulated vendor builds its household graph from the same snapshot without the NCID. That makes the vendor exact on the build date and isolates what happens afterward. We developed the definitions, policies and predictors on 20 randomly drawn counties. We test the primary hypotheses on 60 others, the predictors on the remaining 20 and on the November 2026 file, and transfer on the voter files of Ohio and Washington (section 4.4). Section 2 reviews related work, section 3 defines currency and its decomposition, section 4 gives the data and design, sections 5 to 7 report the measurement, section 8 the predictors and section 9 the other registers, and section 10 discusses implications and limits.

\hypertarget{related-work}{%
\needspace{4\baselineskip}
\section{Related work}\label{related-work}}

Linkage research scores a link on the day it is made. Automated linkers of historical censuses trace a frontier between match rate and false links: Bailey et al.~{[}8{]} find that common methods produce many false links and argue against maximizing match rate, and Abramitzky et al.~{[}1{]} show that conservative rules raise the precision of person links. Health-data linkage separates the linkage rate from linkage quality and shows that linked subsets are selected {[}38{]}. Cluster-level scores {[}2, 7, 64{]}, critiques of F1 {[}21, 34{]} and evaluation categories for group linkage {[}71{]} all ask whether a link or a cluster is right when it is formed. We ask whether the group around a correctly linked person is still right years later.

Currency is a long-standing data-quality dimension {[}9, 10, 12, 13, 81, 94{]}. Currency metrics give the probability that a stored value is still valid given its age. That probability has been taken from an exponential decline with an external rate {[}44{]} and from a lifetime distribution fitted to the data {[}42{]}. It has been conditioned on supplemental data known when the value was stored {[}41{]} and evaluated against observed validity {[}43{]}. Zak and Even {[}96{]} model the real-world state behind a stored value as a continuous-time Markov chain, estimate currency from the fitted transitions and choose when to reacquire the data.

Related work determines which of several values is current {[}27, 28{]}, sets update frequencies {[}83{]}, traces source lag and copying {[}25{]}, cleans stale values {[}66{]} and treats corporate householding as a data-quality task {[}62{]}. The currency metrics above score a stored value of one entity. A household assignment is a set of entities around one of them, and it can go stale while every person-level link stays right. Our scoring by validity at the scoring date follows earlier work on event-aligned accounting {[}14{]}, and our bounds on unscored exits are the worst-case bounds for a missing binary outcome {[}63{]}.

Temporal linkage models how one entity's attributes evolve so that its records link across time {[}15, 16, 20, 33, 59, 60{]}, and collective and group linkage resolve co-occurring references jointly {[}11, 77{]}. Households have been linked across census waves by graph matching and temporal clustering {[}31, 32, 55, 70{]}.

Closest to this paper, Christen et al.~{[}22{]} link households across six census waves, classify how each evolves (preserved, split, merged, moved) and count households preserved over 10 to 50 years. In that count a household is preserved when a one-to-one group link keeps at least two of its members, the count is household-weighted, and it inherits linkage error. What this paper adds is the client's side: exact sets scored against a persistent-identifier reference at sub-annual resolution, together with what a system that stores the set returns. North Carolina voter files are a standard benchmark for person-level linkage and duplicate detection {[}19, 52, 78{]}. Those benchmarks label persons. We label registered co-residence at each of 40 distinct snapshots from 2008 to 2026.

Researchers and statistical offices store households too. Households are built from co-address administrative records {[}36, 88{]}, from anonymized address keys in health records {[}50, 85{]} and from property identifiers assigned at a chosen event date {[}35, 37{]}. Cuccaro-Alamin et al.~{[}24{]} review the strategies, and statistics offices build administrative rosters at census scale {[}53, 67, 68, 87{]}. Such a key is the administrative counterpart of a vendor's household file: it is formed at one date and used at another. Administrative address changes lag real moves {[}56{]}, and record-level addresses in a health register disagree with the census for a sizable share of people {[}30{]}, so a register measures registered co-residence.

Co-registration at an address is an established household proxy in political science {[}45{]}. Voter files have been used to validate surveys, to link persons and to track recorded changes {[}3, 4, 26, 54{]}, and registration responds to moves with a lag that varies with age {[}5{]}. Pew Research Center {[}80{]} matched a national panel to five commercial voter files; match rates ranged from 50\% to 79\%, and only 14\% of panelists who had moved in the past year were found on all five. Household fields in consumer files {[}79{]}, third-party audience segments {[}72, 73{]} and offline broker attributes {[}93{]} are often wrong. Commercial residential histories have been validated against self-reports {[}47{]}, and IP-to-household assignments against operator records {[}23{]}. Each of these studies validates a record at one date. None follows a stored assignment as its evidence ages.

Household composition change has been measured directly. Richards et al.~{[}84{]} fit a competing-risks hazard model of transitions among household types in a national panel. Hays et al.~{[}39{]} find that 35\% of households in a panel of older adults in five North Carolina counties changed composition over four years. Raley et al.~{[}82{]} count composition changes between birth and 18 for children in a national panel, from the change in the set of persistent co-member identifiers between waves of its household roster. That is the exact-set comparison used here; they apply it to survey rosters, and we apply it to a stored assignment. Linked annual UK registers that combine the electoral roll with consumer data give 73.7\% of households unchanged in composition after one year {[}57{]}, a small-area household change index {[}58{]} and estimates of residential moves {[}91{]}. Matching residents by name, as those register studies do, mixes change with linkage error.

Demographic projection models household structure through transition rates {[}92, 97{]} and validates projected totals against a later census, so it does not ask whether one identified household is still intact. Section 8.2 uses multistate Markov models for panel data with interval-censored transitions {[}49, 51, 89{]}. Supplement I gives the literature search.

\hypertarget{measuring-the-currency-of-a-household-assignment}{%
\needspace{7\baselineskip}
\section{Measuring the currency of a household assignment}\label{measuring-the-currency-of-a-household-assignment}}

\hypertarget{anchors-and-currency}{%
\needspace{4\baselineskip}
\subsection{Anchors and currency}\label{anchors-and-currency}}

A system builds its evidence at date \(t\) and answers a query at \(s = t + k\). For a client with anchor type X (address A or person P), the \textbf{currency} \(κ_{X}\)(k) is the probability that the household recorded around the anchor at \(t\) is exactly the household at \(s\), with the client excluded from both sets. For a client \(i\) with returned set \(R_i\) and true set \(T_i\), write \(R'_i = R_i \setminus \{i\}\) and \(T'_i = T_i \setminus \{i\}\); the outcome is

\[E_i = \mathbf{1}\bigl[R'_i = T'_i\bigr]. \qquad (1)\]

We count a client recorded alone at \(t\) whose true household at \(s\) is also empty as current: ``lives alone'' is a correct statement about household composition, even when the client has moved. Section 5.3 reports these clients separately. The convention fits systems that return a set of co-members; a system that returns a persistent household key would need its own rule for a client who has moved out and lives alone. A returned member absent from the rolls at \(s\) breaks the match.

The address anchor's match rate \(P_{A}\), called \textbf{persistence} below, is the share of clients whose current address held a registrant at \(t\). We assume the client brings a correct current address, so \(P_{A}\) is a best case for address resolution. The person anchor is found by construction here, because the client already holds its own build-time identity. Both the households and the reference come from one register: the currency we measure is that of a register-derived household, scored against the same register's later listings. Unless a table says otherwise, \(κ_{A}\) and \(κ_{P}\) are scored among clients still listed at \(s\), \(κ_{A}\) on those whose address is found.

\hypertarget{three-outcomes-per-client}{%
\needspace{4\baselineskip}
\subsection{Three outcomes per client}\label{three-outcomes-per-client}}

Each client gets one of three outcomes: the right household, a wrong household, or no answer. \textbf{Validity} is the exact-household share among answered clients; the whole-file view gives the share of all clients in each outcome (section 7.4). We keep the three apart. Scores that average the first two outcomes with partial credit, or score no answer as the person alone, mix them, and two common conventions (self-inclusive scoring and scoring an unmatched record 0) can reverse which policy looks better (supplement G).

\hypertarget{stayers-and-movers}{%
\needspace{4\baselineskip}
\subsection{Stayers and movers}\label{stayers-and-movers}}

We split clients at \(s\) into \textbf{stayers}, whose current address maps to their own household at \(t\), and \textbf{movers}, everyone else still listed. ``Maps'' is the address re-match a system can make (the address re-match rule): the client's address as recorded at \(s\), matched into the build graph, returns the client's own build household. A client whose unit is unchanged but whose address is recorded at \(s\) in a form the build graph does not hold (a format change or a county relabel) is a mover here. Section 8.2 also uses a second rule, the voter-unit rule, which follows the reference's voter unit and counts such a client as a stayer. With σ the stayer share, \(κ_{S}\) the stayers' currency and μ the movers' currency, total probability gives

\[\kappa_P(k) = \sigma(k)\,\kappa_S(k) + \bigl(1 - \sigma(k)\bigr)\,\mu(k). \qquad (2)\]

Equation (2) separates residential mobility as the register records it, σ, from change inside households that stay put, \(κ_{S}\). The product \(σκ_{S}\) is the share of clients whose stored household and address are both still current. If μ is roughly constant across builds and ages, the decline of the person anchor comes from σ and \(κ_{S}\). Section 5 shows that it is.

\hypertarget{selection}{%
\needspace{4\baselineskip}
\subsection{Selection}\label{selection}}

A stricter acceptance rule θ answers fewer clients. Write \(M_\theta\) for the clients a rule answers, \(V(k,\theta)\) for the exact share of the households it returns to them and \(\sigma_\theta\) for the share of stayers among them. \(\kappa_S(k \mid M_\theta)\) and \(\mu(k \mid M_\theta)\) are the exact shares of the returned households among answered stayers and movers, so they include the effect of wrong person links (section 7.2). Total probability within \(M_\theta\) gives

\[V(k, \theta) = \sigma_\theta\,\kappa_S(k \mid M_\theta) + \bigl(1 - \sigma_\theta\bigr)\,\mu(k \mid M_\theta). \qquad (3)\]

Two conditions turn equation (3) into a bound. Condition (i): among stayers, whether the household changed is independent of being answered, and the returned household is the client's own, so \(\kappa_S(k \mid M_\theta) = \kappa_S(k)\). Condition (ii): the households returned to matched movers are exact no more often than stayers' households, \(\mu(k \mid M_\theta) \le \kappa_S(k)\). Under both, \(V(k,\theta) \le \kappa_S(k)\), with equality when \(\sigma_\theta = 1\). A rule that requires the client's name at the same unit makes \(\sigma_\theta\) close to 1, so V approaches \(κ_{S}\): thresholds plateau. Condition (i) holds only approximately in practice, because a surname changed at marriage or divorce lowers the name score and marks a household that changed. The bound applies within any population a rule answers. A rule that answers only clients of one household type therefore approaches the stayers' currency of that type, which can exceed \(κ_{S}\) pooled over all types. Section 7 measures both.

Against the requirements Heinrich et al.~{[}40{]} state for a data-quality metric, currency has these properties. It is a probability with minimum 0 and maximum 1, and it is interval-scaled. Its configuration is five stated choices: the age of the evidence, the exact-match rule, the lone-mover rule, the rule that a returned member absent at \(s\) breaks the match, and the rule for which clients are trackable. We report the alternative to each except the trackable rule, whose exclusions are counted in supplement E. It aggregates from client to file as a person-weighted mean and, through equation (2), from stayers and movers to all clients. What it needs is a longitudinal reference, which a system rarely has. Section 8 therefore tests predictors that a system can apply from its own build-time features, the supplemental-data design of Heinrich and Klier {[}41{]} applied to a group-valued target.

\hypertarget{data-and-study-design}{%
\needspace{7\baselineskip}
\section{Data and study design}\label{data-and-study-design}}

\hypertarget{data}{%
\needspace{4\baselineskip}
\subsection{Data}\label{data}}

The North Carolina State Board of Elections publishes voter registration snapshots with one 90-column schema, including active, inactive and removed registrants, with age but not birthdate and no household field {[}75{]}. The NCID is filled from November 2008 on. We use 41 snapshot files from November 2008 to November 2026: 9 even-year Election Days, 18 January 1 files, 8 odd-year Novembers, five more 2019 files (two of which are the same extract) and the November 2026 file, so 40 extracts are distinct. The statewide voter-history file gives the build-election vote used in section 6.2, and North Carolina's public address points (NC OneMap) the check of the unit key there. A build is named by its year. The November 2026 file supplies the ten-year values, the evaluation of the predictors on a later file, and the corrected references of section 6.1 that need a file after January 2026. Section 9 uses the voter files of Ohio (July 2017 and October 2026) and Washington (June 2017, June 2018 and October 2026), whose identifiers persist over nine years and which list active and inactive registrants only. Michigan's re-issued identifiers leave no usable pair, and we report no Michigan result (section 9.1). All name and address fields are uppercased before any comparison.

\hypertarget{persons-households-and-the-client-file}{%
\needspace{4\baselineskip}
\subsection{Persons, households and the client file}\label{persons-households-and-the-client-file}}

A \textbf{person} is an NCID. A \textbf{registrant} at \(s\) has an active or inactive row at \(s\); registrants are scored, and the rest of the client file at \(s\) are \textbf{exits}. A person with more than one active-or-inactive row at the build is not a client; at a later snapshot, the handful of registrants with more than one such row are scored on the smallest of their household keys and stay in household sets. The North Carolina file records no age from 1 to 17; a registrant under 18 or with no birth date shows age 0. A person is not a client in two cases, which together exclude every person under 18 at the build. In the first, the age recorded at the build is implausible (0 or above 110) and the nearest snapshot with a plausible age puts them under 18 at the build. In the second, the recorded age is 0 and no snapshot gives a plausible age. Such persons still count as members of other clients' households. A \textbf{household} at \(s\) is the set of registrants who share a voter unit at \(s\), keyed on the register's residential address normalized to a building and a unit; a county relabel of an address is mapped to one canonical building, following residents by NCID, so the relabel is not a move. This mapping applies to the reference only; a graph built from one snapshot cannot make it (section 6.1). A household is \textbf{trackable} at the build when its address is tagged as a likely single-family address or a unit in a multi-unit building and it holds one to five registrants. The \textbf{client file} at build \(t\) is every eligible person in a trackable household at \(t\), followed to every later snapshot wherever they go.

The trackable universe holds 95.0\% of the 3.79 million registrants in the 60 confirmation counties at the 2016 build; the rest are in larger households, group quarters, multi-unit buildings whose record carries no unit, and mobile-home parks and large buildings. We chose the rule on statewide stay rates (supplement C), so it excludes, by design, housing whose membership churns most. The client file also leaves out nonregistrants and is older than the adult population (25.8\% of clients still listed two years after the build were aged 18 to 34 at the build, against 29.2\% of North Carolina adults). The exclusions, the older age mix and the absence of nonregistrants push measured currency up, and counting a member who reaches voting age at the address as a household change pushes it down. Section 6.2 sizes those it can, and the direction of the net selection effect is not identified.

\hypertarget{the-simulated-vendor-and-its-policies}{%
\needspace{4\baselineskip}
\subsection{The simulated vendor and its policies}\label{the-simulated-vendor-and-its-policies}}

At a build date \(t\) the vendor sees every registrant's name and address statewide and forms households by the voter-unit key. A client brings its name and address at \(s\) and asks for its household. The vendor never sees the NCID. Because vendor and reference are built from the same snapshot with the same key, the vendor is exact on the day it builds. Every client is in the build graph, so the match rates we report are not comparable to commercial match rates, which include people a graph never saw. The policies are:

\begin{itemize}
\tightlist
\item
  \textbf{Address anchor (never-rebuilt graph):} re-matches the client's current address against the graph built at \(t\) and returns that unit's household.
\item
  \textbf{Person anchor (last-known household):} returns the client's own household at \(t\); equivalent to a perfect person link.
\item
  \textbf{Annual rebuild:} re-matches against the latest January 1 graph, about ten months old at each scoring date.
\item
  \textbf{Same-date rebuild:} re-matches against a graph built on the scoring date; used only as a same-date baseline for the reference (section 6).
\item
  \textbf{Probabilistic matcher:} a Fellegi-Sunter model {[}29{]} as implemented in Splink {[}61{]}. Levels are first name (exact, first three letters, Soundex), surname (exact, Soundex), middle initial, suffix, and address (same unit, same building, same street and ZIP, same ZIP). Blocking is the union of same unit, same full name statewide, and same surname and first-name Soundex within a ZIP. Age is withheld, so the matcher uses only the name and address a client query brings. The u probabilities, the chance that a level agrees for records of different persons, come from value frequencies with term-frequency adjustment. The prior comes from the build graph. The m probabilities, the chance that a level agrees for records of the same person, come from expectation-maximization (EM) on the development counties, with three blocking rules each estimating the parameters the others hold fixed. A client links to its best candidate at match weight \(W \geq 0\) (posterior 0.5) by default, and a sweep runs from no threshold to \(W \geq 25\).
\end{itemize}

Two further policies serve only the hypothesis tests of section 5.7 and are defined in supplement A: a rebuild every two years from inputs one snapshot old (the lagged rebuild), and an oracle member tracker, which follows members between builds by NCID and returns the household of the plurality carrier.

\hypertarget{study-design}{%
\needspace{4\baselineskip}
\subsection{Study design}\label{study-design}}

Twenty of North Carolina's 100 counties, drawn at random, are development counties: we built the definitions, policies, matcher training and both predictors on them. Sixty confirmation counties test the primary hypotheses and give the statistics of sections 5 to 7, computed identically on both splits; ``confirmation'' names the county split, and the currency accounting of sections 5 to 7 is descriptive. Twenty held-out counties evaluate the predictors and replicate the accounting; their rows also enter every statewide or 80-county computation. Statewide counts made before the draw (universe sizes, stay rates by candidate trackable rule, list maintenance, NCID persistence and re-issue, readdressing, registration lag and minors) shaped the trackable rule and computed no accuracy statistic by policy (supplement C). Table 1 lists what each part of the data evaluates.

Predictions and criteria for each evaluation were recorded before its computation was run; each is reported with the result it governs, analyses reported without a criterion are descriptive, and supplement B lists every criterion (Table S2). Five evaluations used data whose aggregate outcomes had already been examined. For three, the criteria were set after that: the policy checks on the later file, whose statewide currency had been computed; the refresh analysis; and the model in the other registers, which uses the removal-hazard refit of section 9.2 and whose level and split criteria could be computed from values in hand. The household-change model on the held-out counties is reported from a refit of the same model to the same development data, made after its held-out scores had been seen, so its criteria are not presented as set in advance. The accounting on the held-out counties, which followed the tables' test there, has no criterion and is descriptive. Table S2 marks the four that carry criteria. Confirmation values are primary, and development values are labeled as such.

\needspace{12\baselineskip}

\begin{longtable}[]{@{}
  >{\raggedright\arraybackslash}p{(\columnwidth - 4\tabcolsep) * \real{0.1770}}
  >{\raggedright\arraybackslash}p{(\columnwidth - 4\tabcolsep) * \real{0.3717}}
  >{\raggedright\arraybackslash}p{(\columnwidth - 4\tabcolsep) * \real{0.4513}}@{}}
\caption{Parts of the data and what each evaluates.}\tabularnewline
\toprule\noalign{}
\begin{minipage}[b]{\linewidth}\raggedright\hspace{0pt}Part
\end{minipage} & \begin{minipage}[b]{\linewidth}\raggedright\hspace{0pt}Data
\end{minipage} & \begin{minipage}[b]{\linewidth}\raggedright\hspace{0pt}Evaluates
\end{minipage} \\
\midrule\noalign{}
\endfirsthead
\toprule\noalign{}
\begin{minipage}[b]{\linewidth}\raggedright\hspace{0pt}Part
\end{minipage} & \begin{minipage}[b]{\linewidth}\raggedright\hspace{0pt}Data
\end{minipage} & \begin{minipage}[b]{\linewidth}\raggedright\hspace{0pt}Evaluates
\end{minipage} \\
\midrule\noalign{}
\endhead
\bottomrule\noalign{}
\endlastfoot
Development & 20 North Carolina counties, all snapshots & definitions, policies, matcher training, fits of both predictors \\
\arrayrulecolor{black!25}\specialrule{0.25pt}{1.5pt}{1.5pt}\arrayrulecolor{black}
Confirmation & 60 counties & primary hypotheses (section 5.7); the accounting and policy comparisons (sections 5 to 7) \\
\arrayrulecolor{black!25}\specialrule{0.25pt}{1.5pt}{1.5pt}\arrayrulecolor{black}
Held-out & 20 counties & currency tables and household-change model (section 8); replication of the accounting (section 5.6) \\
\arrayrulecolor{black!25}\specialrule{0.25pt}{1.5pt}{1.5pt}\arrayrulecolor{black}
Later file & November 2026: all 100 counties, and the 80 outside development, for the tables; the 60 confirmation counties for ten-year currency, the 2018 build at eight years and the policy checks & tables at a later date (section 8.1); ten-year currency; the 2018 build at eight years (section 5.7); policy checks (section 7.3) \\
\arrayrulecolor{black!25}\specialrule{0.25pt}{1.5pt}{1.5pt}\arrayrulecolor{black}
Refresh & 80 counties outside development, decision at Election Day 2024 & choosing what to refresh (section 8.3) \\
\arrayrulecolor{black!25}\specialrule{0.25pt}{1.5pt}{1.5pt}\arrayrulecolor{black}
Statewide & North Carolina, 2018 build & build-error experiment (section 6.4) \\
\arrayrulecolor{black!25}\specialrule{0.25pt}{1.5pt}{1.5pt}\arrayrulecolor{black}
Other registers & Ohio, July 2017 to October 2026; Washington, June 2017 and June 2018 to October 2026, and June 2017 to June 2018 & transfer of the accounting and the predictors (section 9) \\
\end{longtable}

\hypertarget{inference}{%
\needspace{4\baselineskip}
\subsection{Inference}\label{inference}}

Intervals are 99\% county-cluster percentile bootstrap intervals (2,000 resamples), with the person's county at the build as the cluster, for every decision statistic and for every interval shown unless it is labeled otherwise. They describe variation over counties treated as a draw from a population of counties, with no finite-population correction, and they are nominal. The largest county holds 18.6\% of the confirmation build-date client file and the person-weighted effective number of counties is 16.5 of 60 (development 39.3\% and 5.3 of 20), so development intervals are weak evidence. Supplement B adds a delete-one-county jackknife check for the primary hypotheses, and section 8.2 one for the model comparisons. Differences and ranges in the text are computed from full-precision outputs.

\hypertarget{the-reference-as-a-measurement-model}{%
\needspace{4\baselineskip}
\subsection{The reference as a measurement model}\label{the-reference-as-a-measurement-model}}

The reference records registered co-residence, the household as the register lists it. We treat it as a measurement that errs in known ways:

\begin{itemize}
\tightlist
\item
  \textbf{Stale listings.} A person who left an address can stay listed there, as active or inactive, until the register records the move or removes the person. When the stored household still contains that person, it looks right when it is wrong. When the person arrived after the build and left before \(s\), a stored household that is physically right looks wrong. Exact-set scoring is therefore not monotone in stale listings. Section 6.1 reads them five ways.
\item
  \textbf{Lag, re-issue and unit keys.} Changes are recorded after they happen, persons return under a new NCID, and unit numbers are missing or inconsistent. Section 6.2 measures each.
\item
  \textbf{Unrecorded moves.} A client's move the register never records can leave σ too high, and an unrecorded move of a co-member or an unrecorded arrival can leave \(κ_{S}\) too high. Section 6.3 stresses σ only.
\end{itemize}

\hypertarget{how-fast-stored-assignments-go-stale}{%
\needspace{21\baselineskip}
\section{How fast stored assignments go stale}\label{how-fast-stored-assignments-go-stale}}

\hypertarget{persistence-and-currency}{%
\needspace{18\baselineskip}
\subsection{Persistence and currency}\label{persistence-and-currency}}

\needspace{15\baselineskip}

\begin{longtable}[]{@{}
  >{\raggedright\arraybackslash}p{(\columnwidth - 14\tabcolsep) * \real{0.0460}}
  >{\raggedright\arraybackslash}p{(\columnwidth - 14\tabcolsep) * \real{0.1034}}
  >{\raggedright\arraybackslash}p{(\columnwidth - 14\tabcolsep) * \real{0.1034}}
  >{\raggedright\arraybackslash}p{(\columnwidth - 14\tabcolsep) * \real{0.2184}}
  >{\raggedright\arraybackslash}p{(\columnwidth - 14\tabcolsep) * \real{0.1034}}
  >{\raggedright\arraybackslash}p{(\columnwidth - 14\tabcolsep) * \real{0.1034}}
  >{\raggedright\arraybackslash}p{(\columnwidth - 14\tabcolsep) * \real{0.2184}}
  >{\raggedright\arraybackslash}p{(\columnwidth - 14\tabcolsep) * \real{0.1034}}@{}}
\caption{Persistence and currency, 2016 build, confirmation, registrants at \(s\); \(k = 10\) is scored on the November 2026 file. \(P_{A}\) is the address match rate; \(κ_{A}\) the exact share among address matches; \(κ_{P}\) the exact share of the last-known household; σ, \(κ_{S}\) and μ the stayer share, stayer currency and mover currency; \(σκ_{S}\) the share whose household and address are both current. Brackets are 99\% intervals.}\tabularnewline
\toprule\noalign{}
\begin{minipage}[b]{\linewidth}\raggedright\hspace{0pt}k
\end{minipage} & \begin{minipage}[b]{\linewidth}\raggedright\hspace{0pt}\(P_{A}\)
\end{minipage} & \begin{minipage}[b]{\linewidth}\raggedright\hspace{0pt}\(κ_{A}\)
\end{minipage} & \begin{minipage}[b]{\linewidth}\raggedright\hspace{0pt}\(κ_{P}\) {[}99\%{]}
\end{minipage} & \begin{minipage}[b]{\linewidth}\raggedright\hspace{0pt}σ
\end{minipage} & \begin{minipage}[b]{\linewidth}\raggedright\hspace{0pt}\(κ_{S}\)
\end{minipage} & \begin{minipage}[b]{\linewidth}\raggedright\hspace{0pt}μ {[}99\%{]}
\end{minipage} & \begin{minipage}[b]{\linewidth}\raggedright\hspace{0pt}\(σκ_{S}\)
\end{minipage} \\
\midrule\noalign{}
\endfirsthead
\toprule\noalign{}
\begin{minipage}[b]{\linewidth}\raggedright\hspace{0pt}k
\end{minipage} & \begin{minipage}[b]{\linewidth}\raggedright\hspace{0pt}\(P_{A}\)
\end{minipage} & \begin{minipage}[b]{\linewidth}\raggedright\hspace{0pt}\(κ_{A}\)
\end{minipage} & \begin{minipage}[b]{\linewidth}\raggedright\hspace{0pt}\(κ_{P}\) {[}99\%{]}
\end{minipage} & \begin{minipage}[b]{\linewidth}\raggedright\hspace{0pt}σ
\end{minipage} & \begin{minipage}[b]{\linewidth}\raggedright\hspace{0pt}\(κ_{S}\)
\end{minipage} & \begin{minipage}[b]{\linewidth}\raggedright\hspace{0pt}μ {[}99\%{]}
\end{minipage} & \begin{minipage}[b]{\linewidth}\raggedright\hspace{0pt}\(σκ_{S}\)
\end{minipage} \\
\midrule\noalign{}
\endhead
\bottomrule\noalign{}
\endlastfoot
2 & 0.954 & 0.701 & 0.671 {[}0.654, 0.695{]} & 0.884 & 0.732 & 0.203 {[}0.179, 0.222{]} & 0.647 \\
\arrayrulecolor{black!25}\specialrule{0.25pt}{1.5pt}{1.5pt}\arrayrulecolor{black}
4 & 0.896 & 0.464 & 0.449 {[}0.436, 0.469{]} & 0.740 & 0.535 & 0.205 {[}0.187, 0.221{]} & 0.396 \\
\arrayrulecolor{black!25}\specialrule{0.25pt}{1.5pt}{1.5pt}\arrayrulecolor{black}
6 & 0.863 & 0.380 & 0.379 {[}0.367, 0.397{]} & 0.665 & 0.469 & 0.200 {[}0.185, 0.214{]} & 0.312 \\
\arrayrulecolor{black!25}\specialrule{0.25pt}{1.5pt}{1.5pt}\arrayrulecolor{black}
8 & 0.831 & 0.322 & 0.329 {[}0.319, 0.344{]} & 0.600 & 0.422 & 0.189 {[}0.175, 0.202{]} & 0.254 \\
\arrayrulecolor{black!25}\specialrule{0.25pt}{1.5pt}{1.5pt}\arrayrulecolor{black}
10 & 0.806 & 0.285 & 0.300 {[}0.291, 0.313{]} & 0.554 & 0.393 & 0.185 {[}0.172, 0.197{]} & 0.218 \\
\end{longtable}

Persistence falls slowly and currency falls fast (Table 2, Figure 1). For the 2016 build the stored household is exactly right for fewer than half of the clients still listed by four years, and for about a third by eight. Against the register, \(κ_{P}\) falls by 42\% to 53\% between two and eight years across the five builds (51\% for 2016); section 6 shows how much of the two-year level is the register's own staleness. Only 0.254 of clients at eight years have both the stored household and the stored address still right.

\begin{figure}
\centering
\includegraphics{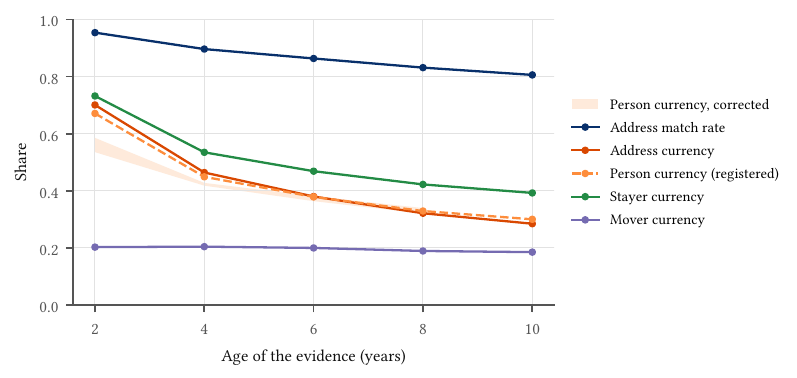}
\Description{Line chart against evidence age of 2 to 10 years, 2016 build: address match rate falls from about 0.95 to 0.81; address currency, person currency and stayer currency fall from about 0.67 to 0.73 at two years to 0.28 to 0.39 at ten; mover currency stays flat near 0.2; a shaded band mostly below the person currency line shows its range under five corrections for stale listings, from two to eight years.}
\caption{Address match rate and the currency of each anchor, stayers and movers by evidence age, 2016 build, with \(k = 10\) scored on the November 2026 file; the shaded band is the person anchor's currency under the five stale-listing corrections of section 6.1 (\(k = 2\) to 8; the two-election lookahead to \(k = 6\)). Confirmation counties.}
\end{figure}

The two anchors give nearly the same currency: across the 20 build and evidence-age cells \(κ_{A}\) − \(κ_{P}\) lies between −0.008 and 0.035. The agreement comes from the address anchor's denominator. For a mover the address anchor returns whoever was recorded at the mover's new address, which is the right household for 0.059 of matched movers at eight years, against μ = 0.189 for the person anchor. The currencies agree because the address anchor answers fewer movers: a mover whose new address held no registrant at \(t\) is unmatched and leaves its denominator.

The decline continues past eight years, more slowly. At ten years \(κ_{P}\) is 0.300, 0.029 {[}0.027, 0.032{]} below eight years. By a midpoint decomposition of equation (2), the stayers' currency carries 0.017 of that fall, the stayer share 0.010 and the mover currency 0.002; μ falls by 0.004 {[}0.002, 0.006{]}.

\hypertarget{the-mover-currency}{%
\needspace{4\baselineskip}
\subsection{The mover currency}\label{the-mover-currency}}

The mover currency hardly changes with the build or with the age of the evidence. Across the five even-year Election Day builds from 2008 to 2016 and the four even evidence ages, μ lies between 0.180 and 0.209 on confirmation (development 0.168 to 0.208): a mover's last-known household is exactly right about one time in five in every build and evidence-age cell. The 99\% half-widths of these cells are 0.008 to 0.045, so the data do not separate a constant μ from one that drifts by a few points, but they rule out a mover currency that falls with age as σ and \(κ_{S}\) do. By years since the recorded move, and for a fixed cohort of early movers, μ stays near one in five. The exception is moves recorded in the first year after the build (0.112), where a late re-registration leaves the build household already stale (supplement D).

The decline of the person anchor therefore comes from the other two terms. For the 2016 build σ falls from 0.884 to 0.600 and \(κ_{S}\) from 0.732 to 0.422 between two and eight years, as members leave, die, age in or arrive around a client who stays. By a midpoint decomposition of equation (2), of the fall of 0.342 in \(κ_{P}\) from two to eight years the stayers' currency carries 0.230, the stayer share 0.108 and the mover currency 0.003. Because the other terms fall and μ does not, movers' stored households are a growing share of what is right: (1 − σ)μ / \(κ_{P}\) is 23.0\% of exact outcomes at eight years.

\hypertarget{movers-who-lived-alone}{%
\needspace{4\baselineskip}
\subsection{Movers who lived alone}\label{movers-who-lived-alone}}

About half of that one in five comes from movers who lived alone. A mover who lived alone at \(t\) and still lives alone at \(s\) counts as current, so lone movers contribute to μ. They are 22\% to 28\% of movers across the 20 cells and supply 41\% to 55\% of μ. Within each size the mover currency stays in a narrow band: 0.336 to 0.407 for movers recorded alone and 0.124 to 0.157 for movers recorded with others (development 0.332 to 0.397 and 0.108 to 0.136; Table S7 gives the 2016 build by household size). So μ mixes two narrow terms, and the first is high because the lone-mover rule counts a lone mover who still lives alone as current.

\hypertarget{build-to-build}{%
\needspace{4\baselineskip}
\subsection{Build to build}\label{build-to-build}}

The level moves from build to build, and it moves with the election calendar. At two years the stayer share is 0.916, 0.867, 0.908, 0.842 and 0.884 for builds 2008 to 2016, swinging with election parity, and \(κ_{P}\) is 0.731, 0.634, 0.710, 0.601 and 0.671 (Table S5). American Community Survey (ACS) mover rates show neither the swing nor a change in mobility of that size (section 6.3), so the pattern is a property of the register. The presidential-year two-year currency falls across 2008, 2012 and 2016 under the registered reference, and the fall steepens under the stale-listing corrections (Table S8). The swing shrinks under the one-election lookahead of section 6.1: two-year \(κ_{P}\) is 0.706, 0.610, 0.663, 0.575 and 0.585 for builds 2008 to 2016.

\hypertarget{the-first-two-years}{%
\needspace{4\baselineskip}
\subsection{The first two years}\label{the-first-two-years}}

The loss is fastest right after the build. We score the last-known household of each even-year build from 2008 to 2016 at up to four dates: the January 1 after the build, the next November file, the January 1 after that and the two-year Election Day. That gives four observations in the first two years for each build except 2010, which has no November 2011 file (Table S18). Two months after the build \(κ_{P}\) is 0.922 to 0.956, at one year 0.805 to 0.831 and at two years 0.601 to 0.731. The first two months lose 0.022 to 0.044 a month, two to four times the rate over the rest of the first year (0.011 to 0.014 a month). The second year loses more when it ends in a presidential election: from fourteen months to two years \(κ_{P}\) falls by 0.183 and 0.184 for the 2010 and 2014 builds, against 0.082 to 0.119 for the 2008, 2012 and 2016 builds, the parity swing of section 5.4. Part of the early loss can be build-date lag (section 6.2). One reading, not tested here, is that the register records a move when the person re-registers, which happens in bursts around elections, so the timing of the loss would also reflect when the register records moves.

\hypertarget{held-out-counties}{%
\needspace{4\baselineskip}
\subsection{Held-out counties}\label{held-out-counties}}

The accounting looks the same on the 20 held-out counties. Across all 20 build and evidence-age cells the held-out values lie within 0.012 of confirmation for \(κ_{P}\), 0.014 for σ, 0.010 for \(κ_{S}\), 0.026 for μ and 0.007 for \(κ_{A}\). For the 2016 build the probabilistic matcher's exact share at \(W \geq 20\) lies within 0.008 (Table S6), and the split by household size within 0.008. The replication carries no criterion, and differences of this size are well inside the held-out intervals (the 99\% intervals for \(κ_{P}\) are 0.07 to 0.09 wide at 10.4 effective counties), so it would not detect differences of a few points.

\hypertarget{primary-hypotheses}{%
\needspace{4\baselineskip}
\subsection{Primary hypotheses}\label{primary-hypotheses}}

We state the three primary hypotheses, tested in seven parts, with \textbf{co-member F1}, a partial-credit score that for client \(i\) is 1 if \(R'_i\) and \(T'_i\) are both empty, 0 if exactly one is empty, and \(2|R'_i \cap T'_i| / (|R'_i| + |T'_i|)\) otherwise, with an unmatched client scored as alone. The gap G(k) between the address anchor's match rate and its mean co-member F1, among registrants at \(s\), measures how much faster a partial-credit score of currency falls than persistence. The hypotheses predict that household accuracy falls faster than match rate, that an oracle member tracker's remaining loss sits in split households (both defined in supplement A), and that fresher evidence cuts the loss.

On the 2016 build every part is supported, and no decision differs from development (Table S2 gives each part and Table S3 its development value and jackknife interval). The gap is positive at two years (0.130 {[}0.118, 0.139{]}) and widens by 0.225 {[}0.214, 0.234{]} from two to eight years, against a threshold of 0.10. The fall in F1 exceeds twice the fall in match rate by 0.102 {[}0.082, 0.122{]}. The widening reaches the threshold in all five builds from 2008 to 2016, smallest for 2010 (0.172 {[}0.165, 0.179{]}). Movers in split households carry at least 0.631 {[}0.621, 0.641{]} of the oracle tracker's loss, against a threshold of 0.50. A rebuild lagged two years keeps the gap at least 0.125 {[}0.117, 0.129{]} at four to eight years and cuts the loss against the never-rebuilt graph by at least 0.311 {[}0.304, 0.319{]}, against a threshold of 0.25.

The decisions condition on registrants at \(s\), 72.8\% of the file at eight years. Leaving each exit's score free, the widening's envelope {[}0.066, 0.399{]} crosses its threshold (equation (S1), supplement F), so section 7.4 gives the whole-file treatment. Supplement B gives the gap under the acceptance rules and at ten years.

We tested the three parts of the first hypothesis again on the 2018 build, scored on the November 2026 file, and they hold among registrants at \(s\). The gap is 0.193 {[}0.180, 0.200{]} at two years and widens by 0.158 {[}0.148, 0.167{]}. The fall in F1 exceeds twice the fall in match rate by 0.055 {[}0.037, 0.073{]}. These results are sensitive to how exits are scored: with every exit scored 1 the widening is 0.067 {[}0.061, 0.073{]}, below its threshold, and the F1 margin of −0.004 {[}−0.015, 0.009{]} is not supported. On the 2016 build both held under that scoring.

\hypertarget{how-wrong-the-reference-could-be}{%
\needspace{7\baselineskip}
\section{How wrong the reference could be}\label{how-wrong-the-reference-could-be}}

\hypertarget{stale-listings}{%
\needspace{4\baselineskip}
\subsection{Stale listings}\label{stale-listings}}

The register is itself stale, so part of the measured decay may come from the reference and not from the stored household. Five alternative references treat some registrants as absent at \(s\), from their inactive status or from later records. The one-election and two-election lookaheads look ahead one and two Election Days: a member is absent at \(s\) if at the later date the person is removed after inactivity or registered in another county with a registration dated on or before \(s\). The inactive-dropped reading drops members listed as inactive at \(s\). The two client-side readings add the client side to the one-election lookahead and to the inactive-dropped reading. A client the same rule flags is scored as a mover whose true set at \(s\) is taken as empty (the lone-mover convention of section 3.1), so it is exact only if stored alone. These two are scenarios resting on that assumption. Each reading can err: some inactive registrants still live where they are listed, some lookahead removals follow a move after \(s\), and the lookahead readings catch only departures the register later records.

\needspace{18\baselineskip}

\begin{longtable}[]{@{}
  >{\raggedright\arraybackslash}p{(\columnwidth - 12\tabcolsep) * \real{0.2286}}
  >{\raggedright\arraybackslash}p{(\columnwidth - 12\tabcolsep) * \real{0.1286}}
  >{\raggedright\arraybackslash}p{(\columnwidth - 12\tabcolsep) * \real{0.1286}}
  >{\raggedright\arraybackslash}p{(\columnwidth - 12\tabcolsep) * \real{0.1286}}
  >{\raggedright\arraybackslash}p{(\columnwidth - 12\tabcolsep) * \real{0.1286}}
  >{\raggedright\arraybackslash}p{(\columnwidth - 12\tabcolsep) * \real{0.1286}}
  >{\raggedright\arraybackslash}p{(\columnwidth - 12\tabcolsep) * \real{0.1286}}@{}}
\caption{Person-anchor currency and the same-date baseline under the registered reference and five corrected references, confirmation, 2016 build. The same-date rebuild is a graph built on the scoring date from the same register; under a corrected reference its score is the fresh graph's build-date exactness. The two-election lookahead needs a file two Election Days after \(s\) and is blank at eight years.}\tabularnewline
\toprule\noalign{}
\begin{minipage}[b]{\linewidth}\raggedright\hspace{0pt}Reference
\end{minipage} & \begin{minipage}[b]{\linewidth}\raggedright\hspace{0pt}\(κ_{P}\), k = 2
\end{minipage} & \begin{minipage}[b]{\linewidth}\raggedright\hspace{0pt}k = 4
\end{minipage} & \begin{minipage}[b]{\linewidth}\raggedright\hspace{0pt}k = 6
\end{minipage} & \begin{minipage}[b]{\linewidth}\raggedright\hspace{0pt}k = 8
\end{minipage} & \begin{minipage}[b]{\linewidth}\raggedright\hspace{0pt}Same-date, k = 2
\end{minipage} & \begin{minipage}[b]{\linewidth}\raggedright\hspace{0pt}Same-date, k = 8
\end{minipage} \\
\midrule\noalign{}
\endfirsthead
\toprule\noalign{}
\begin{minipage}[b]{\linewidth}\raggedright\hspace{0pt}Reference
\end{minipage} & \begin{minipage}[b]{\linewidth}\raggedright\hspace{0pt}\(κ_{P}\), k = 2
\end{minipage} & \begin{minipage}[b]{\linewidth}\raggedright\hspace{0pt}k = 4
\end{minipage} & \begin{minipage}[b]{\linewidth}\raggedright\hspace{0pt}k = 6
\end{minipage} & \begin{minipage}[b]{\linewidth}\raggedright\hspace{0pt}k = 8
\end{minipage} & \begin{minipage}[b]{\linewidth}\raggedright\hspace{0pt}Same-date, k = 2
\end{minipage} & \begin{minipage}[b]{\linewidth}\raggedright\hspace{0pt}Same-date, k = 8
\end{minipage} \\
\midrule\noalign{}
\endhead
\bottomrule\noalign{}
\endlastfoot
Registered & 0.671 & 0.449 & 0.379 & 0.329 & 0.999 & 1.000 \\
One-election lookahead & 0.585 & 0.421 & 0.364 & 0.321 & 0.832 & 0.835 \\
Two-election lookahead & 0.567 & 0.419 & 0.364 & & 0.798 & \\
Inactive dropped & 0.565 & 0.428 & 0.377 & 0.334 & 0.806 & 0.891 \\
Client-side lookahead & 0.557 & 0.418 & 0.366 & 0.326 & 0.771 & 0.776 \\
Client-side inactive & 0.535 & 0.427 & 0.381 & 0.340 & 0.734 & 0.850 \\
\end{longtable}

The corrections are large at two years and small at eight (Table 3). At two years, for the 2016 build they lower \(κ_{P}\) by 0.085 (one-election lookahead) to 0.135 {[}0.120, 0.149{]} (client-side inactive), and for the other four builds by 0.024 to 0.097 (Table S8). At eight years every variant moves \(κ_{P}\) by −0.017 to +0.013 at every build. The held-out counties show the same scale. The correction falls on σ and \(κ_{S}\), and μ rises: at two years under the client-side lookahead σ falls from 0.884 to 0.755 and \(κ_{S}\) from 0.732 to 0.665, while μ rises from 0.203 to 0.224.

A fresh graph inherits every stale listing at \(s\), while a stored assignment is corrected only for the stale listings it still holds. The same-date rebuild scores 0.999 to 1.000 against the register (just short of 1 because the reference merges the units of a relabeled building, which a graph built from one snapshot cannot do) and 0.734 to 0.891 against the corrected references. Under the one-election lookahead its correction is 0.167 at two years and 0.165 at eight, while the correction to \(κ_{P}\) falls from 0.085 to 0.009. The ratio of the two would measure evidence-age loss only if no stored set were right where the fresh set is wrong, and that case occurs, so we do not use the ratio. The address anchor moves the same way: its eight-year currency is 0.322 against the register and 0.309 under the one-election lookahead.

Three conclusions follow:

\begin{enumerate}
\def\labelenumi{\arabic{enumi}.}
\tightlist
\item
  \textbf{Currency falls with evidence age under every reference.} \(κ_{P}\) falls from two to eight years by 0.342 against the register and by 0.196 to 0.265 under the four corrected references with an eight-year value.
\item
  \textbf{The two-year level depends on the reference; the eight-year level is comparatively insensitive to these readings.} \(κ_{P}\) at two years is 0.671 against the register and 0.535 to 0.585 against the corrected references; at eight years it is 0.321 to 0.340 against the register and the four corrected references with an eight-year value. A stale listing matters early, when a stored household still contains most of the people the register lists; by eight years most stored households are wrong for other reasons.
\item
  \textbf{Comparisons between policies keep their order, and the gaps shrink.} The annual rebuild's advantage over the stored household at eight years is 0.471 against the register and 0.326 to 0.387 against the corrected references. The order of the policies at each evidence age is the same under every reference, apart from two policies already tied under the register at six years (supplement E).
\end{enumerate}

The registered reference stays primary: every criterion in this paper is stated against it, the register's listing at \(s\) is the direct measurement of the target, and each corrected reading can err in both directions. We print headline levels with the one-election lookahead value beside them.

\hypertarget{other-register-sensitivities}{%
\needspace{4\baselineskip}
\subsection{Other register sensitivities}\label{other-register-sensitivities}}

The other register-based sensitivities each move \(κ_{P}\) by less than four points, in both directions (Table S9). Dropping inactive registrants from both the stored and the true set raises \(κ_{P}\) by 0.014 to 0.029; this removes the departed person from the stored household too, so it forgives the error that section 6.1 measures. Scoring the same clients two months after \(s\), a short lookahead that overlaps the one-election lookahead, lowers \(κ_{P}\) by 0.002 to 0.021, most at two years. Restricting to clients who voted in the build election raises \(κ_{P}\) by 0.025 to 0.036, which mixes build-date lag with voters being less mobile. The NCID re-issue crosswalk lowers \(κ_{P}\) by 0.002 to 0.004, and dropping clients in buildings with mixed unit keys raises it by 0.006 to 0.010; the voter-unit key disagrees with North Carolina's public address points in 2.0\% of matched buildings at the 2016 build, buildings that hold 7.6\% of registrants. Reweighting the client file to the adult age mix lowers \(κ_{P}\) by 0.003 to 0.007. On the development counties the terms have the same sign and size.

Two choices in the target definition matter more at long ages. A member who was under 18 at \(t\) and registers at the client's address afterward counts as a change in the client's household; dropping those members from both sets raises \(κ_{P}\) by 0.031 at two years and 0.071 at eight. The register cannot separate children reaching voting age in place from adults who moved in and registered, so this is a sensitivity to the target definition. Treating a returned member absent at \(s\) as unknown, in place of a broken match, raises \(κ_{P}\) at eight years from 0.329 to 0.384 (Table 4). Across the target-definition choices and stresses of sections 6.1 to 6.3, eight-year \(κ_{P}\) ranges from 0.26 to 0.40; the registered value, 0.329, is the primary estimate.

\hypertarget{mobility-the-register-does-not-record}{%
\needspace{4\baselineskip}
\subsection{Mobility the register does not record}\label{mobility-the-register-does-not-record}}

A register sees a move only when the person re-registers, so σ is the stayer share as the register records it. Recorded moves run at about half the rate a survey finds for all adults. Over the ACS years 2010 to 2019 and 2021 to 2024 (no standard 2020 release), recorded address changes among adult registrants run at 0.063 a year (0.070 with moves out of county or state, 0.073 with list maintenance), against 0.138 for North Carolina adults in the ACS (table B07001) {[}90{]}; deaths, 0.010 a year, are not counted as moves. Part of the gap is moves the register records late or never, and part is the difference between registrants and all adults. As a stress scenario, we compound the whole unrecorded rate on stayers and score each such stayer as a mover. That lowers \(κ_{P}\) by 0.056 to 0.067 at every evidence age (to 0.264 to 0.271 at eight years). The scenario overstates the effect, because ACS moves concentrate in repeat movers; it moves only σ, so unrecorded moves of co-members and unrecorded arrivals are covered by no reading here; and it overlaps the stale-listing corrections, so the two are never added. The ACS series shows no parity swing.

\hypertarget{build-error}{%
\needspace{4\baselineskip}
\subsection{Build error}\label{build-error}}

The simulated vendor is exact on its build date; a real vendor builds some households wrongly. We perturb the 2018 statewide build and score it at 1.15 and 7.99 years in three arms:

\begin{itemize}
\tightlist
\item
  \textbf{Lag.} Each person's address comes from the build date (40\%) or from a snapshot 10 to 24 months older; a person not listed there is picked up late or omitted.
\item
  \textbf{Noise.} In 10\% of buildings all units merge into one household, and 2\% of addresses fail to match, leaving the person alone.
\item
  \textbf{Both} lag and noise.
\end{itemize}

Build accuracy, the client-weighted share whose built household is exact on the build date, is 0.785 to 0.949 among clients still listed at 7.99 years (Table S22).

At eight years, build error costs less than its face value. The arms keep 0.905 to 0.978 of the exact build's currency (0.290 to 0.313 against 0.320). That is more than their build accuracy: currency is 1.031 to 1.153 times build accuracy times the exact build's currency, inside the predicted range of 0.95 to 1.25. Stale evidence is right mostly for households that were already stable, so a lagged vendor's correct builds stay right more often: their currency exceeds the exact build's by 0.026 {[}0.024, 0.027{]}, as predicted, while random noise is nearly independent of later change (0.004). Some wrong builds also become right, when a missed joiner leaves or a wrongly split household parts: between 0.061 and 0.103 of wrong builds are right at eight years, above the predicted ceiling of 0.05.

At 1.15 years the cost is larger: the arms keep 0.823 to 0.958 of \(κ_{P}\). For these three error types, the exact-build curve lies above the vendor's in every arm, by 2\% to 10\% at eight years.

\hypertarget{matching-rules-fresh-evidence-and-the-whole-file}{%
\needspace{22\baselineskip}
\section{Matching rules, fresh evidence and the whole file}\label{matching-rules-fresh-evidence-and-the-whole-file}}

\begin{figure}
\centering
\includegraphics{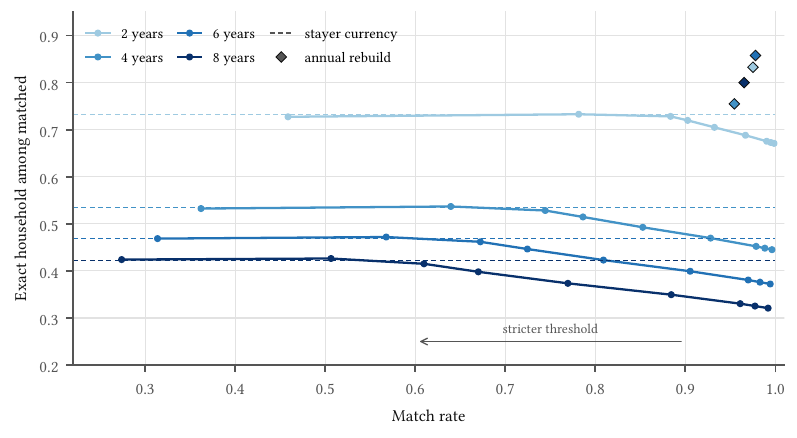}
\Description{Line chart: for each evidence age of 2, 4, 6 and 8 years, the share of matched clients placed in their exact current household falls as the match rate rises with a looser threshold; as the threshold tightens, each curve rises to a plateau near the dashed line for stayer currency at that age, at most slightly above it, and diamonds for annual rebuilds sit well above all curves.}
\caption{Exact household share among matched clients against match rate for the probabilistic matcher's threshold sweep, by age of the evidence; dashed lines are the stayer currency \(κ_{S}\) at each age, diamonds the annual rebuild. Confirmation counties.}
\end{figure}

\needspace{19\baselineskip}

\begin{longtable}[]{@{}
  >{\raggedright\arraybackslash}p{(\columnwidth - 12\tabcolsep) * \real{0.2892}}
  >{\raggedright\arraybackslash}p{(\columnwidth - 12\tabcolsep) * \real{0.0964}}
  >{\raggedright\arraybackslash}p{(\columnwidth - 12\tabcolsep) * \real{0.2169}}
  >{\raggedright\arraybackslash}p{(\columnwidth - 12\tabcolsep) * \real{0.0964}}
  >{\raggedright\arraybackslash}p{(\columnwidth - 12\tabcolsep) * \real{0.0843}}
  >{\raggedright\arraybackslash}p{(\columnwidth - 12\tabcolsep) * \real{0.1205}}
  >{\raggedright\arraybackslash}p{(\columnwidth - 12\tabcolsep) * \real{0.0964}}@{}}
\caption{The validity frontier at \(k = 8\), 2016 build. Exact share is among matched clients (a returned member absent at \(s\) breaks the match), with its 99\% interval; the next column treats such members as unknown. F1 is co-member F1 over all registrants at \(s\) (section 5.7). Person-link precision is the share of probabilistic links that go to the client's own build record.}\tabularnewline
\toprule\noalign{}
\begin{minipage}[b]{\linewidth}\raggedright\hspace{0pt}Rule
\end{minipage} & \begin{minipage}[b]{\linewidth}\raggedright\hspace{0pt}Match rate
\end{minipage} & \begin{minipage}[b]{\linewidth}\raggedright\hspace{0pt}Exact share {[}99\%{]}
\end{minipage} & \begin{minipage}[b]{\linewidth}\raggedright\hspace{0pt}Absent as unknown
\end{minipage} & \begin{minipage}[b]{\linewidth}\raggedright\hspace{0pt}F1
\end{minipage} & \begin{minipage}[b]{\linewidth}\raggedright\hspace{0pt}Person-link precision
\end{minipage} & \begin{minipage}[b]{\linewidth}\raggedright\hspace{0pt}Development exact share
\end{minipage} \\
\midrule\noalign{}
\endfirsthead
\toprule\noalign{}
\begin{minipage}[b]{\linewidth}\raggedright\hspace{0pt}Rule
\end{minipage} & \begin{minipage}[b]{\linewidth}\raggedright\hspace{0pt}Match rate
\end{minipage} & \begin{minipage}[b]{\linewidth}\raggedright\hspace{0pt}Exact share {[}99\%{]}
\end{minipage} & \begin{minipage}[b]{\linewidth}\raggedright\hspace{0pt}Absent as unknown
\end{minipage} & \begin{minipage}[b]{\linewidth}\raggedright\hspace{0pt}F1
\end{minipage} & \begin{minipage}[b]{\linewidth}\raggedright\hspace{0pt}Person-link precision
\end{minipage} & \begin{minipage}[b]{\linewidth}\raggedright\hspace{0pt}Development exact share
\end{minipage} \\
\midrule\noalign{}
\endhead
\bottomrule\noalign{}
\endlastfoot
Probabilistic, no threshold & 0.992 & 0.321 {[}0.310, 0.337{]} & 0.375 & 0.491 & 0.912 & 0.295 \\
Probabilistic, W ≥ 0 (default) & 0.884 & 0.350 {[}0.338, 0.365{]} & 0.407 & 0.499 & 0.978 & 0.323 \\
Probabilistic, W ≥ 10 & 0.670 & 0.398 {[}0.386, 0.414{]} & 0.460 & 0.483 & 0.995 & 0.371 \\
Probabilistic, W ≥ 20 & 0.506 & 0.426 {[}0.413, 0.443{]} & 0.489 & 0.434 & 0.998 & 0.400 \\
Probabilistic, W ≥ 25 & 0.274 & 0.424 {[}0.409, 0.445{]} & 0.484 & 0.328 & 0.998 & 0.396 \\
Address anchor & 0.831 & 0.322 {[}0.307, 0.344{]} & 0.382 & 0.477 & & 0.293 \\
Person anchor & 1.000 & 0.329 {[}0.319, 0.344{]} & 0.384 & 0.503 & & 0.304 \\
Annual rebuild & 0.965 & 0.800 {[}0.790, 0.814{]} & 0.804 & 0.891 & & 0.789 \\
Stayer currency \(κ_{S}\) (no matcher) & & 0.422 {[}0.410, 0.436{]} & & & & 0.397 \\
\end{longtable}

\hypertarget{selection-at-strict-thresholds}{%
\needspace{4\baselineskip}
\subsection{Selection at strict thresholds}\label{selection-at-strict-thresholds}}

Raising the probabilistic matcher's threshold raises the exact share of its matches until it reaches the stayers' currency, and it rises no further (Figure 2). At \(W \geq 20\) the exact share exceeds \(κ_{S}\) by 0.001 to 0.004 at every evidence age (development 0.000 to 0.003), and \(W \geq 25\) sits slightly below \(W \geq 20\) while the match rate halves again. The mechanism is the bound of section 3.4: a strict threshold works by choosing whom to answer. At \(W \geq 20\) and eight years stayers are 0.984 of the matched clients. The matched stayers' currency is 0.430, 0.008 above \(κ_{S}\), against 0.385 for the stayers the rule leaves unmatched, so condition (i) holds only roughly at \(W \geq 20\). The bound that holds is the matched stayers' currency, which sits near pooled \(κ_{S}\) because the rule matches most stayers. Mixed with the matched movers (0.195), the exact share at \(W \geq 20\) is 0.426. At \(W \geq 25\) the matched and unmatched stayers' currencies are 0.426 and 0.420. Historical census linkage sees the same selection when linking rules require household or geographic agreement {[}86{]}, and linked census samples differ systematically from the populations they come from {[}6{]}.

The small excess over pooled \(κ_{S}\) is composition: within clients recorded alone and within clients recorded with others, \(W \geq 20\) lies within −0.003 to +0.002 of that group's \(κ_{S}\) (Table S12, with paired intervals in Table S11). A rule that answers only clients recorded alone is right for 0.497 of them at eight years, above pooled \(κ_{S}\) = 0.422, because it answers a population whose households have less room to change, and among those clients a strict threshold again reaches only their stayers' currency. The price of the threshold is coverage. Moving from \(W \geq 0\) to \(W \geq 20\) on the same client file raises the exact share from 0.350 to 0.426 and cuts the match rate from 0.884 to 0.506; co-member F1 over all clients falls from 0.499 to 0.434, because the clients left unmatched are scored as alone.

\hypertarget{person-link-precision-and-household-validity}{%
\needspace{4\baselineskip}
\subsection{Person-link precision and household validity}\label{person-link-precision-and-household-validity}}

The matcher's wrong person links barely move the result. At its default threshold the probabilistic matcher links 0.978 of its matches to the client's own build record, a precision that reflects linking a register to itself with the same normalization on both sides. Its exact share exceeds \(κ_{P}\) over all clients by 0.017 to 0.021 at every evidence age, and that gap is selection. On the clients it matches, a perfect person link has the same exact share to within 0.0011 at every evidence age on both splits and at both \(W \geq 0\) and \(W \geq 20\) (0.350 at eight years). At \(W \geq 0\) the residual differences change sign with evidence age; at \(W \geq 20\) they stay at about 0.0001 (Table S11).

The wrong links are 2.2\% of matches at eight years, so they could move the exact share by at most that much. They move it by about 0.001 because many of them go to another person at the client's build unit, a relative or a namesake, whose build household is the client's own (on the development counties, most wrong links at two years and about half at eight). Most of the rest are name-only links of clients whose address changed. Errors that cross households can change the exact share by up to their frequency, and the effect depends on which clients are mislinked, so the result describes this matcher's errors.

\hypertarget{fresh-evidence}{%
\needspace{4\baselineskip}
\subsection{Fresh evidence}\label{fresh-evidence}}

Fresher evidence moves currency far more than any threshold does. The annual rebuild returns the exact household for 0.800 of its matches at eight years while matching 0.965 of clients, and for 0.755 to 0.858 across the four scoring dates. Under the one-election lookahead it is 0.711, 0.668, 0.759 and 0.693 at two, four, six and eight years. Supplement G gives household-link survival and support ceilings for methods restricted to build-time candidates.

These comparisons hold on the November 2026 file, at the builds of 2018, 2020, 2022 and 2024, within their limits. \(W \geq 20\) lies within 0.004 of \(κ_{S}\) (point estimate; limit 0.02). The default threshold lies within 0.019 {[}0.018, 0.021{]} of \(κ_{P}\) (limit 0.035), and the address anchor within 0.029 {[}0.027, 0.031{]} of the person anchor (limit 0.05). The annual rebuild scores at least 0.847 {[}0.835, 0.864{]} (floor 0.760). The first and the address-anchor comparison are close to identities, because a full-name match at the same unit is a stayer by construction (Table S14).

\hypertarget{the-whole-client-file}{%
\needspace{4\baselineskip}
\subsection{The whole client file}\label{the-whole-client-file}}

More than a quarter of the 2016 client file has left the rolls by eight years, and exits have no observed household at \(s\). The register records why each person left, so we score each exit by the removal reason on its last record on or before \(s\). A deceased client's right answer is no household, so returning one is wrong. For a client who moved out of state, the returned household is wrong if a returned member is still registered in North Carolina. For a client who moved within the state, it is wrong if a returned member is still at the returned address. For any exit, it is wrong if a returned member is deceased or the returned members are at two or more addresses at \(s\). A client removed after a period of inactivity is unknown in the primary bound, which scores every unknown exit first wrong and then right {[}63{]}, and is read as a probable move in a scenario; other codes are unknown. A policy that returns nothing scores no answer. Supplement F gives the exit accounting by reason (Table S10), and the code release holds the inputs for recomputing it.

\needspace{15\baselineskip}

\begin{longtable}[]{@{}
  >{\raggedright\arraybackslash}p{(\columnwidth - 12\tabcolsep) * \real{0.0526}}
  >{\raggedright\arraybackslash}p{(\columnwidth - 12\tabcolsep) * \real{0.1053}}
  >{\raggedright\arraybackslash}p{(\columnwidth - 12\tabcolsep) * \real{0.2105}}
  >{\raggedright\arraybackslash}p{(\columnwidth - 12\tabcolsep) * \real{0.1316}}
  >{\raggedright\arraybackslash}p{(\columnwidth - 12\tabcolsep) * \real{0.2105}}
  >{\raggedright\arraybackslash}p{(\columnwidth - 12\tabcolsep) * \real{0.1579}}
  >{\raggedright\arraybackslash}p{(\columnwidth - 12\tabcolsep) * \real{0.1316}}@{}}
\caption{Share of the whole confirmation client file by outcome, 2016 build (\(N\) = 3,603,082). The bound for ``right'' runs from scoring every unknown exit wrong to scoring it right; ``upper, inactivity as moves'' is the upper end when those exits are read as moves. Unknown is the person anchor's unknown mass. A bound that scored every exit 0 or 1 would be as wide as the exit share.}\tabularnewline
\toprule\noalign{}
\begin{minipage}[b]{\linewidth}\raggedright\hspace{0pt}k
\end{minipage} & \begin{minipage}[b]{\linewidth}\raggedright\hspace{0pt}Exits
\end{minipage} & \begin{minipage}[b]{\linewidth}\raggedright\hspace{0pt}Person anchor right
\end{minipage} & \begin{minipage}[b]{\linewidth}\raggedright\hspace{0pt}Upper, inactivity as moves
\end{minipage} & \begin{minipage}[b]{\linewidth}\raggedright\hspace{0pt}Address anchor right
\end{minipage} & \begin{minipage}[b]{\linewidth}\raggedright\hspace{0pt}Address anchor, no answer
\end{minipage} & \begin{minipage}[b]{\linewidth}\raggedright\hspace{0pt}Unknown
\end{minipage} \\
\midrule\noalign{}
\endfirsthead
\toprule\noalign{}
\begin{minipage}[b]{\linewidth}\raggedright\hspace{0pt}k
\end{minipage} & \begin{minipage}[b]{\linewidth}\raggedright\hspace{0pt}Exits
\end{minipage} & \begin{minipage}[b]{\linewidth}\raggedright\hspace{0pt}Person anchor right
\end{minipage} & \begin{minipage}[b]{\linewidth}\raggedright\hspace{0pt}Upper, inactivity as moves
\end{minipage} & \begin{minipage}[b]{\linewidth}\raggedright\hspace{0pt}Address anchor right
\end{minipage} & \begin{minipage}[b]{\linewidth}\raggedright\hspace{0pt}Address anchor, no answer
\end{minipage} & \begin{minipage}[b]{\linewidth}\raggedright\hspace{0pt}Unknown
\end{minipage} \\
\midrule\noalign{}
\endhead
\bottomrule\noalign{}
\endlastfoot
2 & 0.061 & {[}0.630, 0.661{]} & 0.650 & {[}0.627, 0.659{]} & 0.043 & 0.031 \\
4 & 0.157 & {[}0.378, 0.471{]} & 0.438 & {[}0.351, 0.442{]} & 0.089 & 0.092 \\
6 & 0.230 & {[}0.292, 0.422{]} & 0.388 & {[}0.253, 0.381{]} & 0.110 & 0.131 \\
8 & 0.272 & {[}0.240, 0.385{]} & 0.355 & {[}0.195, 0.336{]} & 0.132 & 0.146 \\
\end{longtable}

At eight years the person anchor returns the right household for 0.240 to 0.385 of the clients in the whole file (Table 5), and for at most 0.420 if the deceased are excluded rather than scored. Reading removal after inactivity as a move lowers the upper end to 0.355. Identification by reason leaves an unknown mass of 0.146 at eight years, against 0.272 for 0/1 scoring. Unresolved removals after inactivity are 74\% of what remains unknown, and they stay unknown in the primary bound because removal flags for movers are themselves error-prone {[}46{]}. The address anchor answers fewer clients (no answer for 0.043 to 0.132 of the file) and is right for {[}0.195, 0.336{]} at eight years.

\hypertarget{predicting-currency-from-build-time-features}{%
\needspace{4\baselineskip}
\section{Predicting currency from build-time features}\label{predicting-currency-from-build-time-features}}

A system that holds its own build-time features can predict the currency of each stored assignment if currency is a stable function of them. We built two predictors on the development counties and evaluated them on held-out data: lookup tables of currency by build-time cell (section 8.1) and a model of household change from which currency is derived (section 8.2). Section 8.3 asks whether the tables choose better assignments to refresh.

\hypertarget{lookup-tables}{%
\needspace{4\baselineskip}
\subsection{Lookup tables}\label{lookup-tables}}

The target is the person anchor's exact outcome (equation (1)), conditional on the client being listed at \(s\); \(κ_{P}\) times the listed share is the listed-and-exact share of build-date clients (section 9.1). The \textbf{coarse table} has cells by evidence age (2, 4, 6, 8), build, build household size (1 to 5), building type (multi-unit if the address carries a unit designator or the building holds more than one unit) and the client's age at the build in seven bands: 1,400 cells. The \textbf{table with co-member ages} splits each cell further by the number of co-members aged 18 to 25, the mean co-member age in six bands and the client-to-co-member age gap: 21,788 cells. Race, ethnicity, sex and party are not features, and every input is the system's. A cell's value adds the pooled exact share for its evidence age to the cell's count of exact outcomes and divides by the cell's clients plus one, fitted on the 20 development counties over all five builds. A cell absent from the table takes the pooled share at its age (0.641, 0.463, 0.375 and 0.317 at \(k = 2\) to 8), which is also the \textbf{age-only} baseline. For builds the fit never saw, we also fit each table with build parity (presidential or midterm year) in place of build date; the parity form costs 0.0008 in Brier score on development.

We evaluate the tables on the 20 held-out counties and on all 100 counties of the November 2026 file (the later file), against four criteria. The first is a held-out Brier score no more than 0.005 above development. The second is a mean absolute error of the predicted exact share over the 20 build and evidence-age cells (cell error) of at most 0.010. The third is an expected calibration error (ECE) of at most 0.010. The fourth, for the parity tables on the later file, is a cell error of at most 0.030 over four build and evidence-age pairs.

\needspace{28\baselineskip}

\begin{longtable}[]{@{}
  >{\raggedright\arraybackslash}p{(\columnwidth - 14\tabcolsep) * \real{0.1980}}
  >{\raggedright\arraybackslash}p{(\columnwidth - 14\tabcolsep) * \real{0.0693}}
  >{\raggedright\arraybackslash}p{(\columnwidth - 14\tabcolsep) * \real{0.1980}}
  >{\raggedright\arraybackslash}p{(\columnwidth - 14\tabcolsep) * \real{0.1980}}
  >{\raggedright\arraybackslash}p{(\columnwidth - 14\tabcolsep) * \real{0.0891}}
  >{\raggedright\arraybackslash}p{(\columnwidth - 14\tabcolsep) * \real{0.0891}}
  >{\raggedright\arraybackslash}p{(\columnwidth - 14\tabcolsep) * \real{0.0891}}
  >{\raggedright\arraybackslash}p{(\columnwidth - 14\tabcolsep) * \real{0.0693}}@{}}
\caption{The currency tables and two baselines, person anchor: on the 20 held-out counties (18,975,925 client rows: a person contributes one row per build and evidence age; five builds, four ages) and on all 100 counties of the later file at four build and evidence-age pairs (the 80 counties outside development and the 20 held-out counties alone are in Table S16). Age alone is the pooled exact share by evidence age; age × size is a 20-cell table of evidence age by build household size, fitted on the development counties. Skill is 1 minus the predictor's Brier score over the baseline's, with its 99\% interval. Reliability and resolution partition the Brier score {[}69{]} (supplement H); uncertainty is 0.2496 to 0.2499. Cell error is the mean absolute error over the build and evidence-age cells. ECE keeps each block of tied predictions in one bin (supplement H).}\tabularnewline
\toprule\noalign{}
\begin{minipage}[b]{\linewidth}\raggedright\hspace{0pt}Predictor
\end{minipage} & \begin{minipage}[b]{\linewidth}\raggedright\hspace{0pt}Brier
\end{minipage} & \begin{minipage}[b]{\linewidth}\raggedright\hspace{0pt}Skill over age {[}99\%{]}
\end{minipage} & \begin{minipage}[b]{\linewidth}\raggedright\hspace{0pt}Skill over age × size {[}99\%{]}
\end{minipage} & \begin{minipage}[b]{\linewidth}\raggedright\hspace{0pt}Reliability
\end{minipage} & \begin{minipage}[b]{\linewidth}\raggedright\hspace{0pt}Resolution
\end{minipage} & \begin{minipage}[b]{\linewidth}\raggedright\hspace{0pt}Cell error
\end{minipage} & \begin{minipage}[b]{\linewidth}\raggedright\hspace{0pt}ECE
\end{minipage} \\
\midrule\noalign{}
\endfirsthead
\toprule\noalign{}
\begin{minipage}[b]{\linewidth}\raggedright\hspace{0pt}Predictor
\end{minipage} & \begin{minipage}[b]{\linewidth}\raggedright\hspace{0pt}Brier
\end{minipage} & \begin{minipage}[b]{\linewidth}\raggedright\hspace{0pt}Skill over age {[}99\%{]}
\end{minipage} & \begin{minipage}[b]{\linewidth}\raggedright\hspace{0pt}Skill over age × size {[}99\%{]}
\end{minipage} & \begin{minipage}[b]{\linewidth}\raggedright\hspace{0pt}Reliability
\end{minipage} & \begin{minipage}[b]{\linewidth}\raggedright\hspace{0pt}Resolution
\end{minipage} & \begin{minipage}[b]{\linewidth}\raggedright\hspace{0pt}Cell error
\end{minipage} & \begin{minipage}[b]{\linewidth}\raggedright\hspace{0pt}ECE
\end{minipage} \\
\midrule\noalign{}
\endhead
\bottomrule\noalign{}
\endlastfoot
\emph{Held-out counties} & & & & & & & \\*
Coarse table & 0.1976 & 0.159 {[}0.145, 0.180{]} & 0.068 {[}0.059, 0.078{]} & 0.0003 & 0.0526 & 0.0106 & 0.0107 \\
\arrayrulecolor{black!25}\specialrule{0.25pt}{1.5pt}{1.5pt}\arrayrulecolor{black}
Co-member ages & 0.1913 & 0.186 {[}0.172, 0.207{]} & 0.098 {[}0.089, 0.109{]} & 0.0005 & 0.0591 & 0.0090 & 0.0091 \\
\arrayrulecolor{black!25}\specialrule{0.25pt}{1.5pt}{1.5pt}\arrayrulecolor{black}
Age × size & 0.2120 & 0.098 {[}0.088, 0.113{]} & & 0.0007 & 0.0386 & 0.0342 & 0.0236 \\
\arrayrulecolor{black!25}\specialrule{0.25pt}{1.5pt}{1.5pt}\arrayrulecolor{black}
Age alone & 0.2350 & & & 0.0009 & 0.0157 & 0.0377 & 0.0292 \\
\arrayrulecolor{black!25}\specialrule{0.25pt}{1.5pt}{1.5pt}\arrayrulecolor{black}
\emph{Later file, 100 counties} & & & & & & & \\*
Coarse table & 0.1963 & 0.164 {[}0.159, 0.171{]} & 0.061 {[}0.058, 0.065{]} & 0.0011 & 0.0544 & 0.0127 & 0.0122 \\
\arrayrulecolor{black!25}\specialrule{0.25pt}{1.5pt}{1.5pt}\arrayrulecolor{black}
Co-member ages & 0.1896 & 0.193 {[}0.188, 0.199{]} & 0.093 {[}0.090, 0.097{]} & 0.0017 & 0.0617 & 0.0147 & 0.0164 \\
\arrayrulecolor{black!25}\specialrule{0.25pt}{1.5pt}{1.5pt}\arrayrulecolor{black}
Age × size & 0.2091 & 0.109 {[}0.106, 0.115{]} & & 0.0006 & 0.0411 & 0.0153 & 0.0168 \\
\arrayrulecolor{black!25}\specialrule{0.25pt}{1.5pt}{1.5pt}\arrayrulecolor{black}
Age alone & 0.2348 & & & 0.0003 & 0.0151 & 0.0142 & 0.0142 \\
\end{longtable}

On the held-out counties, at the point estimate, the table with co-member ages met all three criteria: its Brier score is 0.0001 below development, its cell error 0.0090 and its ECE 0.0091. The coarse table met the Brier criterion, 0.0004 above development, and missed the other two by under a thousandth (cell error 0.0106, ECE 0.0107). The intervals settle neither result. At 20 counties, 10.4 effective, both tables' intervals for cell error and ECE contain 0.010 (Table S13). On the later file both parity tables met the cell-error criterion (0.0127 and 0.0147).

The tables' gain is resolution. Their skill over age alone is 0.16 for the coarse table and 0.19 with co-member ages, on the held-out counties and on the later file alike; reliability is at most 0.002 for every predictor, while resolution rises from 0.015 to 0.016 for age alone to 0.053 to 0.062 for the tables. Half to two-thirds of the gain comes from household size, by a sequential comparison with the age × size table, and the tables' skill over that table has every 99\% interval clear of zero. A calibration difference between the tables and age alone is not established: age alone's ECE is 0.029 {[}0.003, 0.073{]} on the held-out counties, an interval that overlaps the tables'.

On the held-out counties both tables underpredict every one of the 20 cells, by 0.011 on average for the coarse table and 0.009 with co-member ages. The miss is county-level variation, which the tables do not hold. Development offsets by county have a standard deviation of about 0.024, so the mean offset of a block of 10.4 effective counties has a standard deviation of about 0.007, against a criterion of 0.010.

The tables drift at a later date, most for the newest build. On the later file across all 100 counties, which include the development counties on which the pooled age shares were fitted, age alone scores a cell error of 0.0142, below the table with co-member ages (0.0147). On the 80 counties outside development both tables beat age alone (0.0109 and 0.0130 against 0.0216; Table S16). The largest cell error on the later file is for the 2024 build at two years, overpredicted by 0.037 and 0.043 (age alone misses it by 0.009; Table S15). The parity table pools the presidential builds of 2008, 2012 and 2016, whose two-year currency on the development counties fell from 0.712 to 0.681 to 0.637. On the same counties the 2024 build is lower again (0.625).

The tables are fitted to the registered reference and overpredict a corrected one. On the 20 held-out counties, scored against the one-election lookahead without refitting, the tables overpredict by up to 0.059 and 0.060 at two years, 0.034 and 0.036 at four, 0.025 and 0.027 at six and 0.013 and 0.014 at eight. That is 19 cells: the 40 files before November 2026 hold no Election Day after November 2024, so the 2016 build at eight years has no later Election Day there, and Table 3 takes it from the November 2026 file. Supplement H gives the Brier scores and cell errors against that reading.

\hypertarget{a-model-of-household-change}{%
\needspace{4\baselineskip}
\subsection{A model of household change}\label{a-model-of-household-change}}

A table cell holds a smoothed estimate of the exact share and nothing about why households change. It is silent on evidence ages the fit did not hold and on how the loss divides among σ, \(κ_{S}\) and μ. A model of the events that change a household gives all three, if the currency it derives holds up on counties its fit never read.

The model is a multistate Markov model with hidden states {[}49, 51{]}. At each snapshot it observes two facts about a listed client: whether the household around the client is exactly the build household, and whether the client is at the build unit. \(κ_{P}\), σ, \(κ_{S}\) and μ are their marginal and conditional probabilities. A latent continuous-time chain moves among intact, broken and re-formed households, each at the build unit or elsewhere, crossed with listed and unlisted, because currency is scored only among listed clients. A broken household passes through two latent phases so that its return rate can fall with time {[}89{]}. Event rates depend on the ages of the client and of each co-member, household size, building type, time since the build and calendar year, through 205 coefficients; the table with co-member ages, which encodes the same ages differently, holds 21,788 cells. As in Zak and Even {[}96{]}, currency is derived from fitted transitions, but the validity target here is a set of people.

We fit the model by a composite likelihood over the 195,288,247 consecutive snapshot pairs of the development counties from 2008 to 2026, with transitions out to 122 months after the build. Currency at any age up to ten years is then derived from the fitted rates, inside the fit window, and is not fitted as a target. It is the model's probability of an exact household among clients still listed, carried forward from the build state, with unlisted clients changing household at the listed rates (supplement K gives the states, rates and likelihood).

We selected the model on the development counties and evaluated it on the held-out counties with two criteria. The first is a Brier score below the coarse table's, with the 99\% upper bound of the difference below zero. The second is derived σ, \(κ_{S}\) and μ each within a mean absolute error of 0.02 of the observed values over the 20 build and evidence-age cells, under the voter-unit stayer rule the model is fitted to (section 3.3).

\needspace{18\baselineskip}

\begin{longtable}[]{@{}
  >{\raggedright\arraybackslash}p{(\columnwidth - 8\tabcolsep) * \real{0.2222}}
  >{\raggedright\arraybackslash}p{(\columnwidth - 8\tabcolsep) * \real{0.3077}}
  >{\raggedright\arraybackslash}p{(\columnwidth - 8\tabcolsep) * \real{0.1368}}
  >{\raggedright\arraybackslash}p{(\columnwidth - 8\tabcolsep) * \real{0.0769}}
  >{\raggedright\arraybackslash}p{(\columnwidth - 8\tabcolsep) * \real{0.2564}}@{}}
\caption{The household-change model and the tables on the 20 held-out counties, person anchor, \(k = 2\) to 8, 18,975,925 client rows. Evidence age is an input of every predictor. ``Model minus predictor'' is the difference in Brier score with its 99\% county-cluster interval; negative values favor the model. Age × size and age alone are the baselines of Table 6, scored on the same clients, without paired intervals.}\tabularnewline
\toprule\noalign{}
\begin{minipage}[b]{\linewidth}\raggedright\hspace{0pt}Predictor
\end{minipage} & \begin{minipage}[b]{\linewidth}\raggedright\hspace{0pt}Other build-time inputs
\end{minipage} & \begin{minipage}[b]{\linewidth}\raggedright\hspace{0pt}Size
\end{minipage} & \begin{minipage}[b]{\linewidth}\raggedright\hspace{0pt}Brier
\end{minipage} & \begin{minipage}[b]{\linewidth}\raggedright\hspace{0pt}Model minus predictor {[}99\%{]}
\end{minipage} \\
\midrule\noalign{}
\endfirsthead
\toprule\noalign{}
\begin{minipage}[b]{\linewidth}\raggedright\hspace{0pt}Predictor
\end{minipage} & \begin{minipage}[b]{\linewidth}\raggedright\hspace{0pt}Other build-time inputs
\end{minipage} & \begin{minipage}[b]{\linewidth}\raggedright\hspace{0pt}Size
\end{minipage} & \begin{minipage}[b]{\linewidth}\raggedright\hspace{0pt}Brier
\end{minipage} & \begin{minipage}[b]{\linewidth}\raggedright\hspace{0pt}Model minus predictor {[}99\%{]}
\end{minipage} \\
\midrule\noalign{}
\endhead
\bottomrule\noalign{}
\endlastfoot
Household-change model & client age, each co-member's age band, building type, build & 205 coefficients & 0.19631 & \\
\arrayrulecolor{black!25}\specialrule{0.25pt}{1.5pt}{1.5pt}\arrayrulecolor{black}
Table with co-member ages & the same ages, summarized & 21,788 cells & 0.19126 & 0.00505 {[}0.00412, 0.00591{]} \\
\arrayrulecolor{black!25}\specialrule{0.25pt}{1.5pt}{1.5pt}\arrayrulecolor{black}
Coarse table & client age, household size, building type, build & 1,400 cells & 0.19760 & −0.00129 {[}−0.00198, −0.00076{]} \\
\arrayrulecolor{black!25}\specialrule{0.25pt}{1.5pt}{1.5pt}\arrayrulecolor{black}
Coarse table without building type & client age, household size, build & 700 cells & 0.19925 & −0.00294 {[}−0.00414, −0.00189{]} \\
\arrayrulecolor{black!25}\specialrule{0.25pt}{1.5pt}{1.5pt}\arrayrulecolor{black}
Age × size & household size & 20 cells & 0.21201 & \\
\arrayrulecolor{black!25}\specialrule{0.25pt}{1.5pt}{1.5pt}\arrayrulecolor{black}
Age alone & none & 4 cells & 0.23501 & \\
\end{longtable}

The model beats the coarse table and loses to the table that sees the same ages. Its held-out Brier score is 0.19631 (Table 7), against 0.19760 for the coarse table, a difference of −0.00129 {[}−0.00198, −0.00076{]}, and its squared error is lower in every one of the 20 counties, by 0.00022 to 0.00269 per client (one-sided sign test, p below one in a million; a delete-one-county jackknife-t interval is {[}−0.00214, −0.00045{]}). Its skill over age alone is 0.165, between the coarse table's 0.159 and the finer table's 0.186, and its skill over age × size is 0.074. The coarse table is the weaker comparator, because the model also sees each co-member's age band. The closer comparison is the table with co-member ages, which uses the same age information in another encoding and beats the model by 0.00505 {[}0.00412, 0.00591{]} (jackknife-t {[}0.00387, 0.00623{]}), four times the model's margin over the coarse table, and in every county.

The derived split meets its criterion only under the voter-unit rule, with mean absolute errors of 0.0137 for σ, 0.0093 for \(κ_{S}\) and 0.0187 for μ. Against the split of section 3.3, which every other decomposition in the paper uses, they are 0.0126, 0.0090 and 0.0309: the derived mover currency misses the measured one by 0.031, above the limit of 0.02 (supplement K.5). The derived \(κ_{P}\) also runs low. It is below the observed value in 19 of the 20 cells, by 0.014 on average, so its cell error is 0.0138 against 0.0106 for the coarse table. At ten years, an age the tables do not hold, it is low by 0.007 to 0.017 for the 2008 to 2014 builds.

The model needs 205 coefficients where the finer table needs 21,788 cells, and it gives currency at any evidence age inside the fit window. Its split of the loss is not validated under the paper's stayer rule.

\hypertarget{choosing-what-to-refresh}{%
\needspace{4\baselineskip}
\subsection{Choosing what to refresh}\label{choosing-what-to-refresh}}

A currency estimate should inform which stored assignments to refresh when the budget covers only part of the file. Choosing what to refresh under a budget is the freshness problem studied in {[}17, 83, 96{]}. At the decision date, Election Day 2024, the stored file holds, for each person, the household of the 2018, 2020 or 2022 build, chosen by a fixed hash among the builds at which the person has a row. Refreshing a person's assignment replaces it with the household the 2024 graph returns, and every assignment is scored in November 2026. Each rule ranks the file and refreshes its first fraction b: at random, oldest build first, by predicted gain, by lowest predicted currency (a naive rule) or by the realized gain (an oracle). The predicted gain is the currency a predictor gives the person if refreshed now (two-year-old evidence at scoring, the same build-time features) minus the currency it gives the stored household; the predictors are age × size and the two parity tables. The criterion for each table, against age × size, had three parts. The gain in the exact share at \(b = 0.25\) must have a lower 99\% bound of at least 0.002. The upper 99\% bounds at \(b = 0.10\) and 0.50 must be at least zero. The point gain at \(b = 0.25\) must be positive on the persons who hold all three builds.

\begin{longtable}[]{@{}
  >{\raggedright\arraybackslash}p{(\columnwidth - 6\tabcolsep) * \real{0.1159}}
  >{\raggedright\arraybackslash}p{(\columnwidth - 6\tabcolsep) * \real{0.3478}}
  >{\raggedright\arraybackslash}p{(\columnwidth - 6\tabcolsep) * \real{0.1159}}
  >{\raggedright\arraybackslash}p{(\columnwidth - 6\tabcolsep) * \real{0.4203}}@{}}
\caption{The refresh analysis at Election Day 2024, scored in November 2026: 80 counties outside development, cost one per person, 4,898,857 persons in the file. Exact is the share of the 4,485,983 persons still listed at the scoring date whose household is exactly right after the first fraction b of the file (the budget) is refreshed in the rule's order; it is 0.4126 with no refresh and 0.6695 with all refreshed. D is the exact share minus that under age × size at the same budget, with its 99\% county-cluster interval. Age × size and the two tables rank by predicted gain; lowest predicted first ranks by the stored currency the table with co-member ages predicts; the oracle ranks by the realized gain and is the ceiling.}\tabularnewline
\toprule\noalign{}
\begin{minipage}[b]{\linewidth}\raggedright\hspace{0pt}Budget
\end{minipage} & \begin{minipage}[b]{\linewidth}\raggedright\hspace{0pt}Rule
\end{minipage} & \begin{minipage}[b]{\linewidth}\raggedright\hspace{0pt}Exact
\end{minipage} & \begin{minipage}[b]{\linewidth}\raggedright\hspace{0pt}D {[}99\%{]}
\end{minipage} \\
\midrule\noalign{}
\endfirsthead
\toprule\noalign{}
\begin{minipage}[b]{\linewidth}\raggedright\hspace{0pt}Budget
\end{minipage} & \begin{minipage}[b]{\linewidth}\raggedright\hspace{0pt}Rule
\end{minipage} & \begin{minipage}[b]{\linewidth}\raggedright\hspace{0pt}Exact
\end{minipage} & \begin{minipage}[b]{\linewidth}\raggedright\hspace{0pt}D {[}99\%{]}
\end{minipage} \\
\midrule\noalign{}
\endhead
\bottomrule\noalign{}
\endlastfoot
10\% & Random & 0.4382 & −0.0262 {[}−0.0273, −0.0246{]} \\*
& Oldest first & 0.4488 & −0.0157 {[}−0.0167, −0.0144{]} \\*
& Age × size & 0.4644 & \\*
& Coarse table & 0.4631 & −0.0013 {[}−0.0023, −0.00003{]} \\*
& Co-member ages & 0.4626 & −0.0019 {[}−0.0035, 0.0002{]} \\*
& Lowest predicted first & 0.4579 & −0.0065 {[}−0.0104, −0.0034{]} \\*
& Oracle & 0.5218 & 0.0574 {[}0.0563, 0.0584{]} \\
25\% & Random & 0.4768 & −0.0426 {[}−0.0449, −0.0395{]} \\*
& Oldest first & 0.5029 & −0.0165 {[}−0.0187, −0.0137{]} \\*
& Age × size & 0.5194 & \\*
& Coarse table & 0.5184 & −0.0010 {[}−0.0029, 0.0010{]} \\*
& Co-member ages & 0.5199 & 0.0005 {[}−0.0015, 0.0029{]} \\*
& Lowest predicted first & 0.5156 & −0.0038 {[}−0.0079, −0.00033{]} \\*
& Oracle & 0.6856 & 0.1662 {[}0.1637, 0.1683{]} \\
50\% & Random & 0.5411 & −0.0366 {[}−0.0382, −0.0348{]} \\*
& Oldest first & 0.5695 & −0.0081 {[}−0.0088, −0.0074{]} \\*
& Age × size & 0.5776 & \\*
& Coarse table & 0.5851 & 0.0075 {[}0.0035, 0.0111{]} \\*
& Co-member ages & 0.5887 & 0.0110 {[}0.0076, 0.0140{]} \\*
& Lowest predicted first & 0.5931 & 0.0155 {[}0.0107, 0.0193{]} \\*
& Oracle & 0.6907 & 0.1130 {[}0.1103, 0.1153{]} \\
\end{longtable}

Neither table met its criterion (Table 8). At \(b = 0.25\) the table with co-member ages gains 0.0005 {[}−0.0015, 0.0029{]} over age × size and the coarse table loses 0.0010 {[}−0.0029, 0.0010{]}; on the complete cases, the persons with rows at all three builds, both lose (−0.0013 and −0.0039). At \(b = 0.10\) both lose, by 0.0019 and 0.0013; the interval of the table with co-member ages contains zero, and that of the coarse table ends just below it (−0.00003). At \(b = 0.50\) both gain, by 0.0110 and 0.0075 with intervals clear of zero, but the naive rule gains more there (0.0155) and loses at the smaller budgets. Age × size, the simplest predicted-gain rule, beats oldest first by 0.0165 {[}0.0137, 0.0187{]} at \(b = 0.25\). Every rule stays far from the oracle (0.6856 against 0.5194).

The result does not depend on the stratum or the date. On all 100 counties and on the complete cases neither table gains at \(b = 0.25\) with an interval clear of zero, and both gain at \(b = 0.50\). At a second decision date, Election Day 2022, both tables lose at \(b = 0.10\) with intervals below zero, neither gains at \(b = 0.25\) with an interval clear of zero, and both gain at \(b = 0.50\) (supplement J, Table S17). The predicted gain is the difference of two predictions from one table, the refreshed and the stored, both made from the stored build-time features, and nothing in it is fitted to the gain itself; a model fitted to the gain is left open (section 10.3).

\hypertarget{two-other-registers}{%
\needspace{7\baselineskip}
\section{Two other registers}\label{two-other-registers}}

\hypertarget{the-accounting}{%
\needspace{4\baselineskip}
\subsection{The accounting}\label{the-accounting}}

Ohio and Washington keep identifiers that persist over nine years. Michigan re-issued many of its identifiers, and the name and birth-year crosswalk that would link its long gaps missed its bias criterion in Ohio and Washington (−0.032 and −0.021, against a limit of 0.02; Table S2), so we report no Michigan result. The Ohio and Washington lists hold active and inactive registrants only, so exits cannot be classified by reason. As in North Carolina, clients under 18 on the build date are excluded and remain only as co-members. The pairs are Ohio from July 2017 to October 2026 (9.26 years) and Washington from June 2017 and June 2018 to October 2026 (9.25 and 8.25 years), with a one-year Washington pair. The criteria come from North Carolina statewide values. The North Carolina fit is the least-squares line of statewide currency against the gap over five North Carolina build pairs scored in November 2026 (supplement L.2). \(κ_{P}\) must lie within 0.05 of that fit at the same gap. The state's own stayer share combined with North Carolina's \(κ_{S}\) and μ must give \(κ_{P}\) within 0.025. And μ must lie inside a band of 0.15 to 0.25.

Washington's level and decomposition meet the stated tolerances against the North Carolina fit; Ohio's do not. In Washington, at 9.25 years (June 2017 to October 2026) and 8.25 years (June 2018 to October 2026), \(κ_{P}\) is 0.314 and 0.325, within 0.007 of the North Carolina fit, and the decomposition holds, with deviations of 0.014 and 0.013. In Ohio, at 9.26 years, \(κ_{P}\) is 0.377 {[}0.361, 0.397{]}, 0.071 above the fit, and Ohio's higher stayer share explains only 0.019 of that gap (deviation 0.051). In both states μ is inside the band at the long gaps (0.242 in Ohio, 0.212 and 0.207 in Washington), and Washington's one-year pair, at 0.253, falls just outside it. The orderings \(κ_{S}\) \textgreater{} \(κ_{P}\) \textgreater{} μ, and higher currency for clients recorded alone than for those recorded with others, hold everywhere, and the two anchors agree within 0.032 (Tables S2 and S20).

Part of Ohio's higher level is who stays on its list. \(κ_{P}\) is conditional on still being listed at \(s\), and Ohio keeps fewer clients listed: 0.615 of its build-date clients at 9.26 years, against 0.718 in Washington and 0.666 to 0.703 in North Carolina for builds of 2016 to 2018 scored in November 2026. The listed-and-exact share of build-date clients, \(κ_{P}\) times the listed share, is 0.232 in Ohio, 0.225 and 0.238 at the two long Washington pairs and 0.196 to 0.225 in North Carolina. It counts every exit as not exact, so it is a lower end, not the currency of the whole file, and it does not order the registers' full-file currency. On it the Ohio gap against Washington falls from 0.063 to 0.007; against North Carolina, interpolating at 9.26 years between the 2018 build at eight years (0.225) and the 2016 build at ten (0.196) gives 0.207, so 0.026 of the gap remains. One reading is that a list that removes more of the clients whose households changed raises the currency of those it keeps; section 9.2 tests one form of it.

\hypertarget{the-predictors-and-the-model}{%
\needspace{4\baselineskip}
\subsection{The predictors and the model}\label{the-predictors-and-the-model}}

Levels need not carry across registers, so we score ranking after recalibration in the large. For each predictor and pair one intercept \(\alpha\) is fitted on the scored pair itself, \(\operatorname{logit}(p') = \alpha + \operatorname{logit}(p)\), where \(p\) is the predictor's probability, \(p'\) the recalibrated one and \(\operatorname{logit}(p) = \log\bigl(p/(1-p)\bigr)\). The comparisons are therefore in-sample for the level. They are made at the three long pairs, with the tables' eight-year cells; a build that is not an even-year November takes the mean of the midterm and presidential cells. The ranking transfers. The table with co-member ages beats age × size by 0.0247 in Ohio and by 0.0181 and 0.0186 in Washington, against a predicted minimum of 0.010 (North Carolina gives 0.022 to 0.023), and every comparison among the tables keeps its North Carolina order with intervals clear of zero (Table S21).

We applied the household-change model of section 8.2 at the same pairs with only the removal hazard of its listing process refitted, to each pair's retention: the share of build-date clients still listed at \(s\), by cell and building type (supplement M). A simpler one-offset version of that refit puts Ohio's removal rate at about 1.4 times North Carolina's and Washington's at about 0.86 at both long pairs. No household outcome entered the refit, only retention at \(s\) on the scored pair, which a system would not have at the build. Its criteria, applied at the three long pairs, were four. \(κ_{P}\) must lie within 0.025 of the observed value in each state, with the Ohio error at most half that without the refit. σ, \(κ_{S}\) and μ must each lie within 0.03. Its raw Brier score must beat each table's with the upper bound below zero. And the Ohio refit must lower raw Brier (Table S2). The Washington one-year pair is reported without a criterion.

The model carries Washington's level and misses Ohio's. In Washington it is within 0.015 and 0.017 of \(κ_{P}\) at the long pairs, inside the limit; at the one-year pair it is 0.036 high and its μ 0.063 low. In Ohio it is 0.054 low with the refit and 0.053 without it (0.3236 and 0.3241 against 0.3773; Table S23), and the refit makes the Ohio Brier score very slightly worse (+0.0001), so the listing process as modeled does not explain Ohio's level.

The split misses in both states. Under the section 3.3 rule σ is off by −0.070 in Ohio and by 0.059 and 0.053 in Washington, μ is 0.041 to 0.060 low and \(κ_{S}\) is within 0.025. Under the voter-unit rule (format changes only in Ohio and Washington; supplement M) σ is off by up to 0.081. Observed μ itself, on the section 3.3 rule, is above every North Carolina value at eight to ten years (0.182 to 0.191) in both states (Table S20). On raw Brier (Table S24) the model beats the coarse table in Washington (−0.0015 {[}−0.0027, −0.0008{]} and −0.0011 {[}−0.0020, −0.0006{]}) but not in Ohio (+0.0004 {[}−0.0002, +0.0013{]}), and it trails the table with co-member ages at every pair, by 0.005 to 0.010. It beats age × size by 0.012 to 0.017 at every long pair, and after the same in-sample intercept shift it beats the coarse table at all three (Ohio −0.0024 {[}−0.0030, −0.0019{]}).

The model therefore does not carry Ohio's level better than the tables, and its split does not transfer. In the fit, a higher removal rate removes clients still recorded at the build address and lowers σ, while Ohio's observed σ is higher. Carrying Ohio's level would need a register-specific rule for who is removed, which retention alone cannot identify.

\hypertarget{discussion}{%
\needspace{7\baselineskip}
\section{Discussion}\label{discussion}}

\hypertarget{what-to-ask-of-a-household-system}{%
\needspace{4\baselineskip}
\subsection{What to ask of a household system}\label{what-to-ask-of-a-household-system}}

If you buy householded data, ask the system three questions, and check the third on your own file. How old is the evidence behind each stored assignment, measured from its last refresh? What does the system return when the client's name and address no longer agree with that evidence? What share of clients does it answer? A match rate answers only the last.

The curves here carry to a system that refreshes members separately only if each assignment's age is measured from its own last refresh and refreshes are not triggered by the change they would detect. Person-link precision measures the person link only (section 7.2), and its small effect here holds for a matcher whose wrong links are few or mostly stay inside the client's household. For the matcher measured here, a stricter threshold buys validity up to about the stayers' currency of the population it answers, at the cost of coverage (section 7.1). An annual rebuild, from evidence about ten months old, is right for 0.800 of its matches at eight years (section 7.3).

If you build such a system, your own build-time features let you state currency, among clients still listed, for segments of your stored households. The calibration error is near a point on held-out counties, and drift is about four points for the newest build at a later date. These are probabilities for segments. They do not say which client's household is still right.

A system that can refresh only part of its file should rank by the gain predicted from evidence age and household size: that rule beat refreshing the oldest first by 0.0165 at a quarter of the file. The tables did not show the required improvement on that ranking at a quarter, their criterion. They gained about a point at half and lost at a tenth at the 2022 decision date (section 8.3). A system using another register's fit should set the level on its own file (section 9).

A researcher who attaches a household key built at one date to outcomes observed years later is in the same position as the buyer. The key still links, but in the register measured here the people it groups are exactly the household for only about a third of clients eight years on. The age of the key belongs in the analysis as a measured source of error, beside the linkage rate.

The same holds one step downstream of the household file. Much household-level targeting and measurement joins two files on a household key: one party holds attributes of the people in a household, and another holds the devices, addresses or accounts that reach it. The join is made on the address, and at eight years the address is still found for 0.831 of clients. Attributes of the people it carries are only as current as the household behind them. An attribute derived from the people recorded at the build date, such as an age band or a life stage, is sure to describe the people now there only when the household is unchanged. That holds for 0.701 of found addresses at two years, and for 75.5\% to 85.8\% even when the evidence is ten months old. An attribute of the dwelling does not age this way. We measured a register-derived graph and no commercial file, so these are rates for registered households. The test itself is one any holder of a household key can run: store the date of the evidence behind each assignment, and report exact-household currency by that age beside the match rate.

\hypertarget{limitations}{%
\needspace{4\baselineskip}
\subsection{Limitations}\label{limitations}}

The reference is who the voter file records at a unit. Nonregistrants are invisible. In a national sample matched to several voter and commercial lists, at least 11\% of adult citizens were on none of them and a further 12\% were mislisted {[}48{]}. Here households look smaller than they are, and a child who registers at 18 looks like a household change.

Reference staleness is a main threat to validity. The stale-listing corrections lower two-year currency by 0.085 to 0.135 on the 2016 build (0.024 to 0.135 across builds) and move eight-year currency by −0.009 to +0.010 (−0.017 to +0.013 across builds; section 6.1). Build-date lag works the other way (section 6.2). Unrecorded moves and arrivals of clients and co-members are incompletely observed, and no analysis here bounds their net effect (section 6.3). The parts overlap and are not added.

Starting from an exact build isolates decay from build-time linkage error. Section 6.4 maps simulated errors; we measured no real vendor's error.

The measurement covers one state from 2008 to 2026 and registrants in trackable households derived from the register. Transfer covers Ohio at one build and Washington at two. Confirmation is an out-of-county replication within the same state and years, and in-movers cross county lines. Inference is nominal (section 4.5). Each policy is a stylized version of one way a system answers from stale evidence. Five evaluations used data whose aggregate outcomes had been examined (section 4.4).

\hypertarget{open-problems}{%
\needspace{4\baselineskip}
\subsection{Open problems}\label{open-problems}}

\begin{enumerate}
\def\labelenumi{\arabic{enumi}.}
\tightlist
\item
  Build-time error of a real vendor: section 6.4 maps simulated errors, and only a vendor file scored against the register would measure one.
\item
  Removal after inactivity: can the register's later records identify where those exits went, and narrow the largest remaining unknown?
\item
  Beating \(κ_{S}\) at scale: a method that returns the subset of people at the person's build and current addresses with the highest expected co-member F1, or abstains, needs its own criteria and a holdout not used in this paper, such as a later file or another state.
\item
  Drift across builds: the 2018 build's widening of the gap from two to eight years (0.158) is the lowest of six, and presidential two-year currency has fallen across the 2008, 2012, 2016 and 2024 builds (0.731 to 0.660 on the confirmation counties, Tables S5 and S14). Whether full-age curves keep shifting is a question for the 2020 to 2024 builds as they reach eight years.
\item
  Level across registers: whether a register-specific rule for which clients are removed would carry Ohio's level, and how to estimate that rule without the outcomes it is meant to predict.
\item
  Refresh by predicted gain: a model fitted to the gain itself, not the difference of two outcome predictions (section 8.3).
\end{enumerate}

\hypertarget{conclusion}{%
\needspace{4\baselineskip}
\section{Conclusion}\label{conclusion}}

A stored household assignment loses validity much faster than its anchor becomes hard to find. In North Carolina's register a 2016 household is still exactly right for 0.322 of the clients whose address is found eight years later (0.329 for the person anchor). That is about a third under every reading of the register's stale listings, and 0.26 to 0.40 across the target alternatives and stresses of section 6, a range for registered co-residence and not a bound on physical households.

A mover's stored household is right about one time in five at every evaluated age from two to eight years, so the loss runs through who stays and how households that stay change. Stricter thresholds of this matcher mainly select stayers, whose households are most likely intact, and fresh evidence moves currency most. Currency can be predicted for segments from a system's own build-time features, and a multistate model of household change derives it at ages up to ten years, on average about 0.014 low over the 20 two-to-eight-year cells on the held-out counties. Both give probabilities for segments and leave open which client's household is still right, and the model did not carry its level to Ohio.

A household file should carry the age of the evidence behind every assignment, and a system that reports how many clients it found should also report how many of the households it returned are still right. Finding the person or the address stays easy for years. The household around them does not stay right: eight years on, about one in three is still exact.

\textbf{Supplementary material.} Supplements A to M are in the supplementary material; Table S1 gives the notation.

\hypertarget{statements}{%
\needspace{4\baselineskip}
\section*{Statements}\label{statements}}
\addcontentsline{toc}{section}{Statements}

\textbf{Data and ethics.} The study uses North Carolina voter registration snapshots that the State Board of Elections publishes as public records under state law {[}74{]}, with the statewide voter-history file and North Carolina's public address points, and, for section 9, voter files of Ohio and Washington, obtained by the author as an independent researcher on each state's published terms: Ohio posts its statewide file for public download {[}76{]}, and Washington furnishes its registration lists to any person who requests them and bars their use for commercial advertising or solicitation {[}95{]}. The study complies with these terms: no file is sold, shared or redistributed, and no record is used to advertise to, solicit or contact anyone. Michigan's file, obtained through the state's qualified voter file data request under its Freedom of Information Act {[}65{]}, was inspected for identifier persistence only and is not otherwise used. The research involved no interaction or intervention with any person and no linkage to any person-level source beyond the state's own registration and voter-history files (the address points carry no person data); names and addresses were used only to build the household key and to simulate matching, inside a private environment, and are never published or redistributed. No individual is identified or contacted, and no result is used to profile or target anyone. Only aggregates are reported or released, over 20 or more counties in North Carolina and statewide in the other states: apart from the county split, which records which county belongs to which sample, the release holds no county-level data and no person-level record, released table cells with fewer than 10 persons are merged, persons coded as minors at the build are excluded as clients (section 4.2), and no analysis by race or ethnicity is reported. The author, an independent researcher without an institutional review board, judged that secondary analysis of public records with no interaction or intervention did not require review, the category the US Common Rule exempts as secondary research on publicly available information (45 CFR § 46.104(d)(4)); no institutional determination was made.

\textbf{Data availability.} The North Carolina snapshots, statewide voter-history file and address points are public and downloadable from the State Board of Elections and NC OneMap; the code release lists each file with its checksum. Ohio and Washington supply current files under the terms above; the files used here are identified by file name, size and SHA-256 in the code release (supplement L.1) and are not redistributed. No person-level data are shared.

\textbf{Code availability.} The code is in Python. It is at {\def\UrlBreaks{\do\/}\def\UrlBigBreaks{}\url{https://github.com/methodtrace/2026-stale-household-assignments}}, public when this paper is posted, and archived at \url{https://doi.org/10.5281/zenodo.24026471}. From the 41 public North Carolina snapshot files, one end-to-end run reproduces every North Carolina number in this paper and its supplement at the precision printed. The exceptions are the build-election voting restriction and the address-point audit (section 6.2 and Table S9), which need the same copies of the weekly voter-history file and of the address-point service; the release records the checksums of the copies used. The Ohio and Washington results need the dated extracts listed with their sizes and SHA-256 checksums in the release. The states supply their current files, and earlier extracts may not be obtainable; with other files the code runs and gives different values. The release holds code, a description of the input schema and aggregate results, including the fitted tables, the household-change model's specification and fitted coefficients, and the county split; it holds no input records. The run fits and scores the unmerged tables; the released table files merge cells with fewer than 10 persons, so scoring from those files alone differs slightly.

\textbf{Funding.} This work received no external funding; the author bore its costs.

\textbf{Competing interests.} The author declares no competing interests. The author is a co-inventor on US patents 10,419,883 and 11,223,926, which cover statistically associating mobile and non-mobile devices with geographic areas: a device's location observations are clustered, each cluster is scored for the likelihood that the device's user lives in a household there, and the device is assigned to the highest-scoring area. No patented method is used in this work.

\textbf{Use of AI tools.} During the preparation of this work the author used Claude Opus (Anthropic) to review the computation code, search the literature, and help with drafting the text. After using these tools, the author reviewed and edited the content as needed and takes full responsibility for the content of the published article.

\end{document}